\documentclass[journal]{IEEEtran}
\usepackage{graphicx}
\usepackage{epstopdf}
\usepackage{subfigure}
\usepackage{setspace}
\usepackage{amsmath}
\usepackage{mleftright}
\begin{document}

\title{Theoretical Analysis of Multi Integrating RX Front-Ends for Lossy Broad-Band Channels}

\author{Antroy Roy Chowdhury, Shovan Maity,~\IEEEmembership{Student Member,~IEEE} and Shreyas Sen,~\IEEEmembership{Senior Member,~IEEE}
}

%\markboth{Manuscript}%

\maketitle

\begin{abstract}
In this paper, we present a theoretical analysis of different integrating front-ends employed in broad-band communications through \textit{lossy} channels. Time-domain receivers for broad-band communication typically deal with large integrated noise due to its high bandwidth of operation. However, unlike traditional wireline systems that are typically not noise-limited, channels with high channel-loss render the input signal swing to be very small imposing several challenges in RX design as the circuits operate in the noise-limited regime. This simultaneous high integrated noise and low signal-swing limits the maximum achievable data-rate for a target bit-error-rate (BER) and deteriorates the energy-efficiency of the RX. In this work, transient, noise and gain performance of different standard signaling blocks have been obtained with closed-form expressions and are validated through spice-simulations. Multi-integrator cascade has been proposed which provides significant gain with relatively lower power consumption than the standard gain elements. Also, maximum achievable data-rate and optimum energy efficiency for different channel losses have been obtained theoretically for different architectures revealing their advantages and limitations. All the pertaining circuits have been designed in 65 nm CMOS process with a 1 V supply voltage. 
\end{abstract}


% \fi
\IEEEpeerreviewmaketitle

\section{Introduction}
\IEEEPARstart{A}{s} different communication standards are emerging, the primary focus of any type of communication remains on the optimization of energy-efficiency, i.e the energy spent on transmitting a single bit, as well as maximizing the data-rate. In wireless communication, due to the practical form-factor of antennas and FCC limitations of usable frequency bands, modulation or frequency up-conversion in the TX followed by a demodulation or frequency down-conversion in the RX are of absolute necessity. Due to deployment of modulation/demodulation schemes and high channel loss ($\sim 60-80$ dB) of wireless channel, the power consumption is sufficiently high in wireless transceivers. In popular wireless techniques such as Wi-Fi, near-field communication (NFC), Zigbee, BTLE etc. the best energy-efficiency that can be achieved is close to few nJ/bit \cite{DAC_16_Sen}. However, in wireline communication through electrical links, availability of a broad-band for transmission and significantly lower channel loss reduce the transceiver energy consumption drastically and increase the data-rate. For typical wireline applications such as backplane, ethernet, USB etc. the energy efficiency can be reduced to as low as $\approx 1-10$ pJ/bit (\cite{Tony_IO, Choi_IO, Mansuri_IO}). A wireline channel being low-pass with a very small low-frequency channel loss, while designing transcievers for wireline links, major emphasis is given in mitigation of inter-symbol interference (ISI) to increase the data-rate. The reason behind this is that the data-rate or speed limitation primarily comes from ISI and not from the integrated noise. Hence, analyzing the noise performance and gain provided by different signaling blocks and their implications on the overall performance of different RX architectures are largely overlooked. However, for applications which utilize broad-band channels with sufficiently large low frequency channel loss (\cite{proximity_ISSCC}-\cite{TBME_HBC}), transmitted signal gets highly attenuated which drastically degrades the BER (bit-error-rate) performance of the RX. Employing broad-band communication in such situation, even though an energy-efficient solution, necessitates the need for analyzing the noise and gain performance of different signaling blocks which essentially limits the maximum achievable data-rate.

In this work, we explore different plausible receiver (RX) architectures that can be employed for broad-band communication through lossy wireline-like channels (with channel loss $>20$ dB over all frequencies, unlike the traditional wireline channels such as FR4 traces or cables). An extensive theoretical analysis of different signaling blocks, such as sampler, integrator and LNA, typically used in low-loss wireline link RXs has been carried out to find their suitability and performance while employed for high-loss applications. A theory for finding closed-form expressions for the input-referred noise of the sampler and integrator has been delineated. Multi-integrator cascades are proposed as low-power gain elements. An accurate expression for the gain of integrators has been derived and extended for multi-integrator cascades. Based on these analyses, optimum performances of different RX architectures are found as a function of the channel loss.

Rest of the paper is organized as follows: Section II provides the motivation behind the work explaining few typical examples of lossy channels which employ broad-band communication. Section III expounds different plausible RX architectures suitable for these kind of channels. Section IV deals with deriving closed-form equations and a rigorous performance analysis of different signaling blocks used in the RX architectures in terms of gain, integrated noise and power consumption. Section V utilizes the analyses of section IV to find out the performance of different RX architectures with different channel loss followed by their comparison in section VI. Section VII concludes the paper.

\section{Motivation: Broad-band communication through lossy wireline-like channels}
Considering the advantages of energy-efficient wireline techniques over wireless, several wireline-like  communication techniques have evolved for short distance communications. These include mm-scale proximity communication (\cite{proximity_ISSCC, proximity}) and meter-scale human-body communication (\cite{HBC_CICC}-\cite{TBME_HBC}). The major difference between the wireline and wireline-like channels arises from the significant loss provided by the wireline-like channels. Following are the descriptions of two major applications where wireline-like channels are being used for data-communication between devices. 
\subsection{mm-Scale Proximity Communication}

\begin{figure}[!t]
\centering
\includegraphics[width=3in]{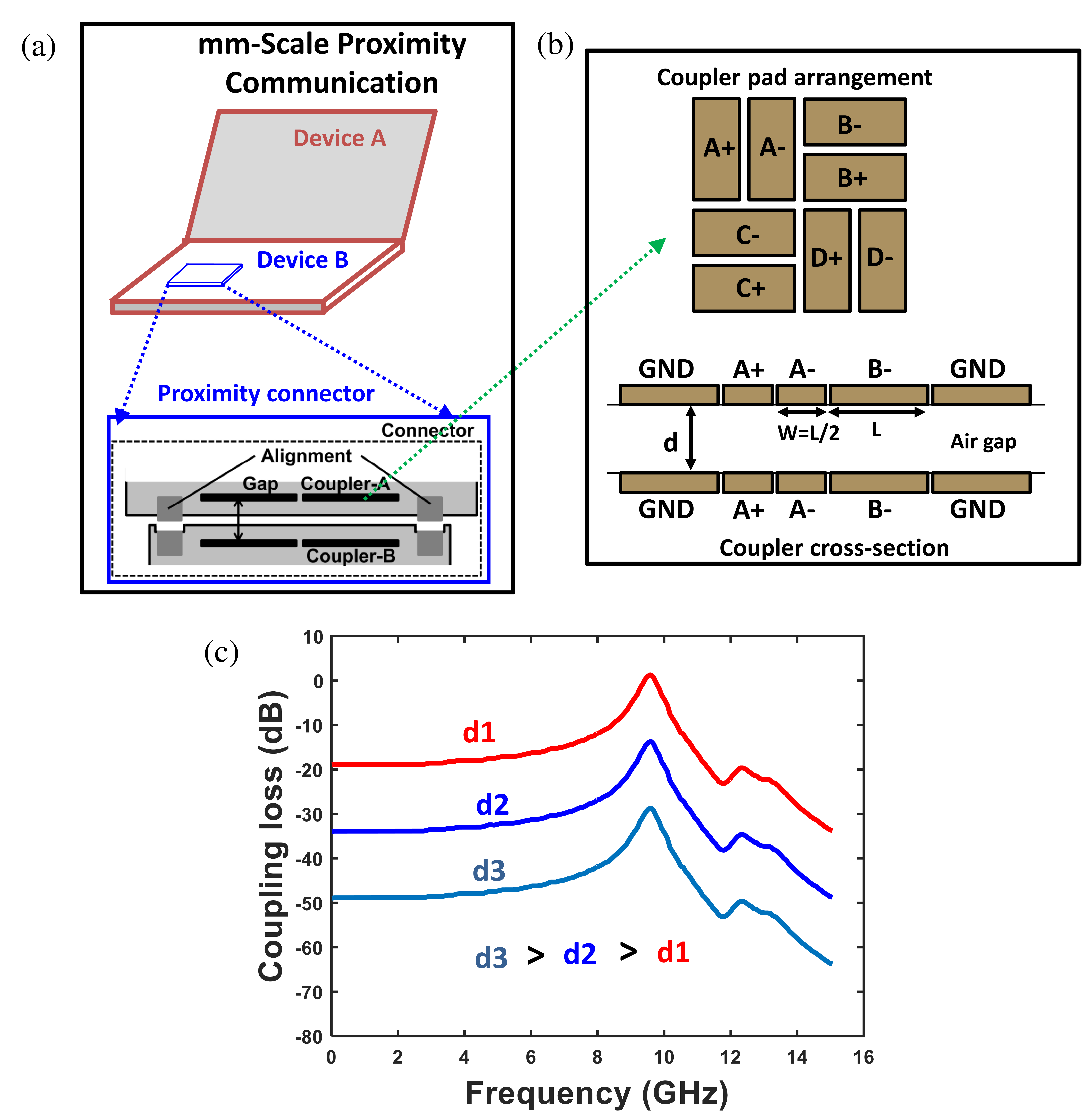}
\caption{mm-scale proximity communication: (a) Communication between two portable devices in close proximity through proximity connector as demonstrated in \cite{proximity}, (b) arrangement of coupler pads for minimizing effect of cross-talk and coupler cross-section, (c) variation of channel loss or coupling loss with frequency for different coupler separations.}
\label{proximity_pic}
\end{figure}

In this type of communication, two devices in close proximity to each other communicate through capacitive coupling as shown in Fig. \ref{proximity_pic}. Here the channel behaves like a simple capacitive divider giving a maximally flat frequency response and hence, a proximity connector can utilize wireline-like baseband signalling and mixed-signal processing for energy-efficient implementation. As the channel behaves as a capacitive divider, the channel loss or coupling loss largely depends on the coupler plate dimensions and the separation between the couplers. For a fixed coupler size, the coupling loss increases with increase in the coupler separation. \cite{proximity_ISSCC} demonstrates a transceiver with energy efficiency $\approx 4$ pJ/bit for $19$ dB coupling loss and a maximum achievable coupler separation of $0.8$ mm satisfying a BER of $10^{-12}$. However, with increase in coupler separation and hence the coupler loss (Fig. \ref{proximity_pic}(c)), the BER increases rapidly as the total noise contributed by different sources become comparable to the signal.  Hence, analysis of different conventional RX architectures, finding their limits of operation as the channel loss increases and exploring new RX architectures suitable for channels with sufficiently high loss find their role pretty important.

\begin{figure}[!t]
\centering
\includegraphics[width=\columnwidth]{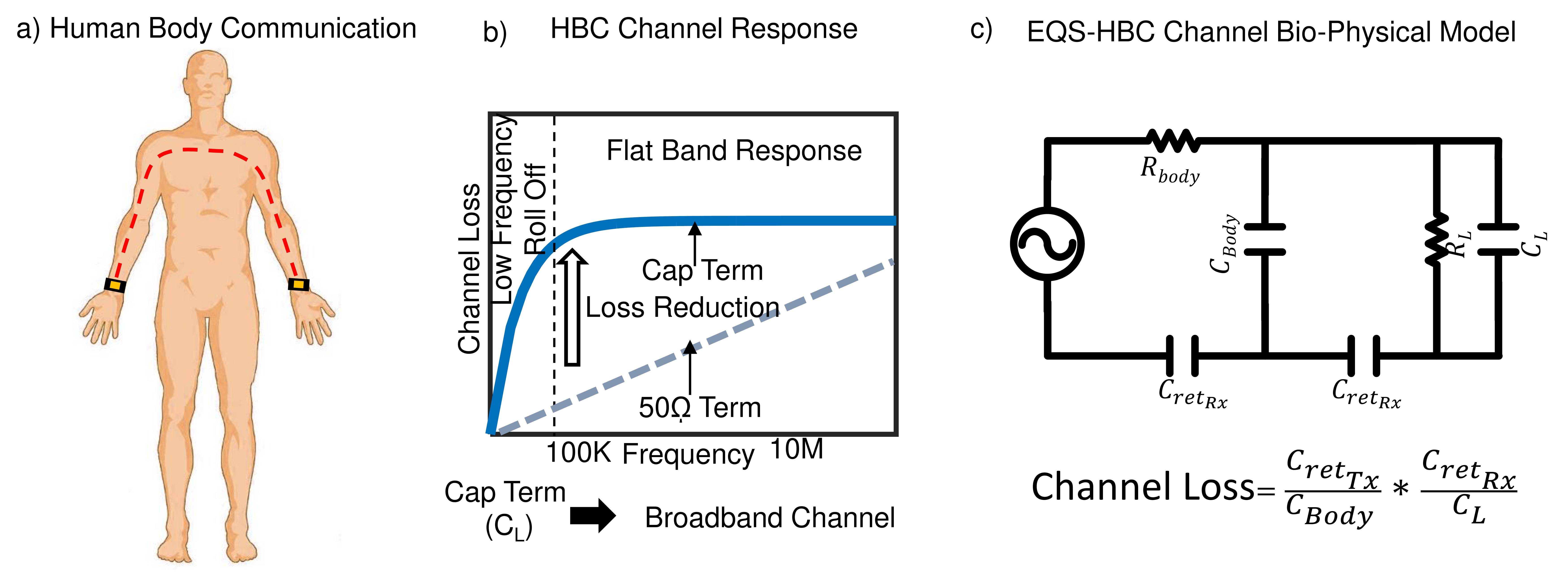}
\caption{Human Body Communication: (a) Communication between two wearable devices using the body as the communication medium. (b) Flatband channel response using high impedance capacitive termination at the receiver end (c) Equivalent circuit model of the HBC channel: $R_{body}$ is the body resistance, $C_{body}$ is the capacitance between the body and earth ground, $C_{retTx}$ , $C_{retRx}$ are the parasitic return path capacitance between the transmitter and receiver respectively, $R_L$, $C_L$ are the load resistance and capacitance respectively.}
\label{hbc_basic}
\end{figure}

\begin{figure}[!t]
\centering
\includegraphics[height=1.4in]{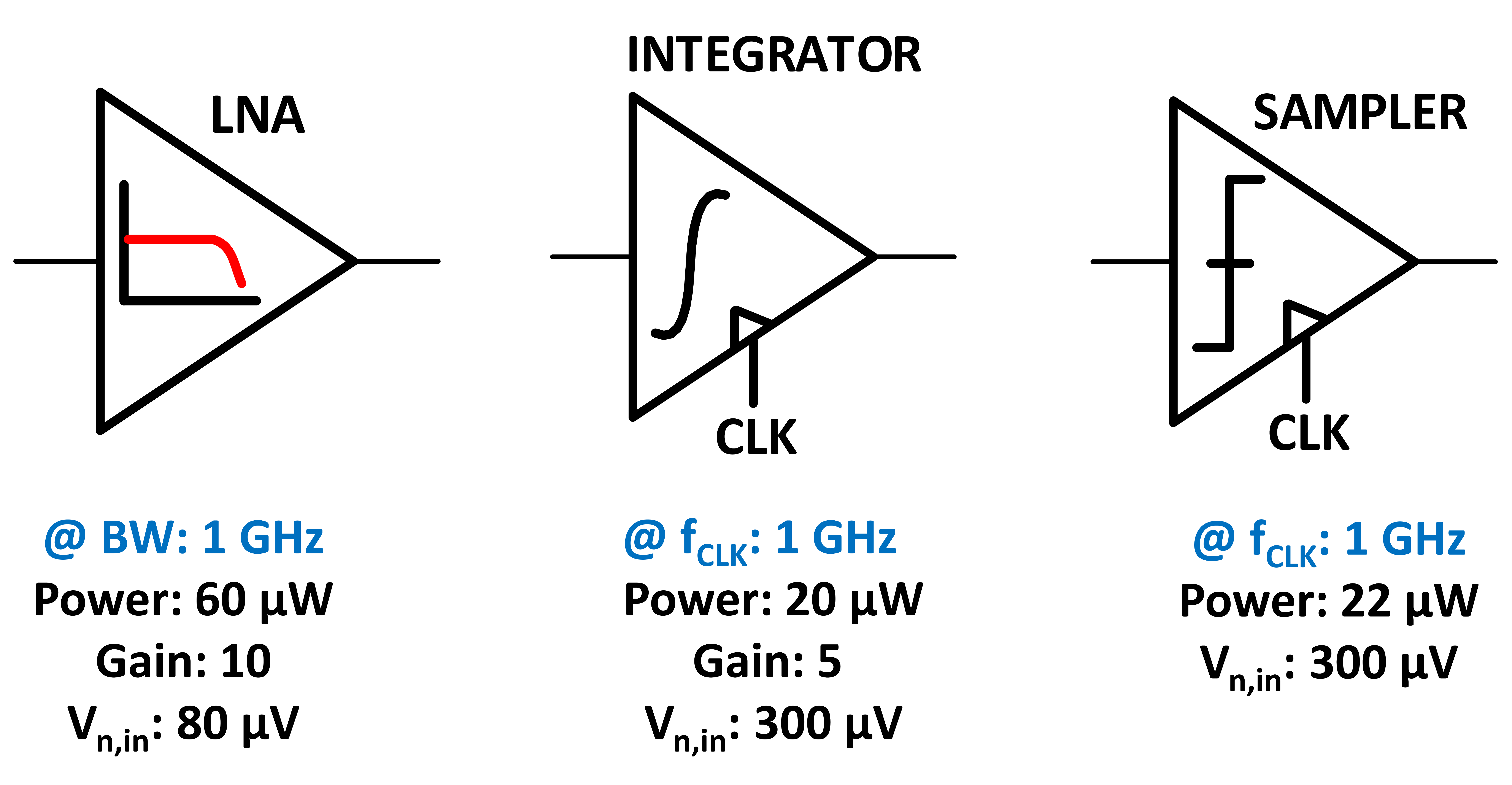}
\caption{Different signaling blocks, required in the RX front-end for NRZ communication through lossy broad-band channels. Typical simulated performance of each block is shown for a $1$ V supply in $65$ nm CMOS technology. The numbers are obtained by optimally sizing the transistors in each block for a band-width of $1$ GHz and using a load capacitor of $10$ fF.}
\label{signaling_blocks}
\end{figure}

\begin{figure*}[!t]
\centering
\includegraphics[width=6.2in]{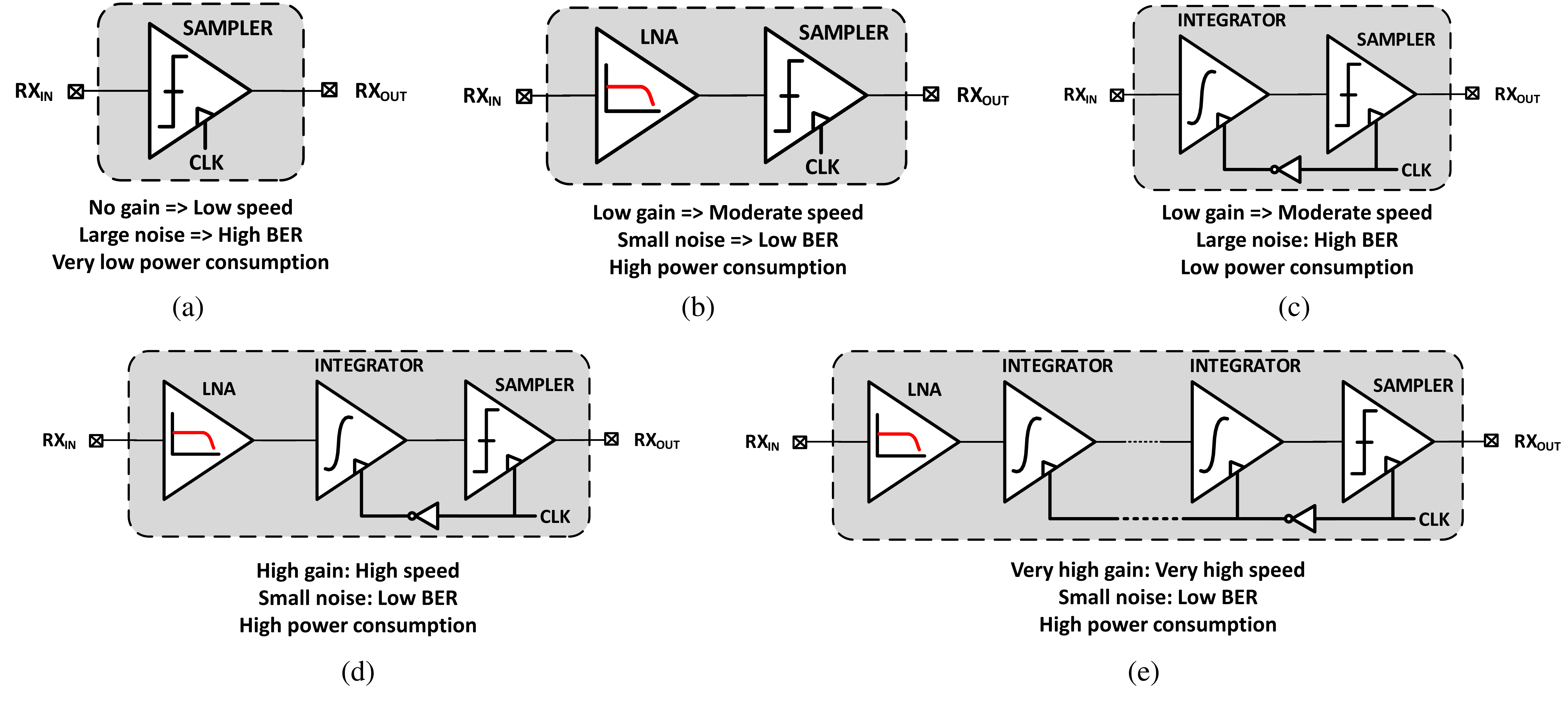}
\caption{Different possible topology of the RX front-end based on signaling blocks shown in Fig. \ref{signaling_blocks}: (a) Only sampler, (b) LNA + sampler, (c) integrator + sampler, (d) LNA + integrator + sampler, (e) LNA + multi-integrator cascade + sampler}
\label{archi}
\end{figure*}

\subsection{Human-Body Communication (HBC)}
Another emerging example of broadband communication where channel loss turns out to be critical is human-body communication shown in Fig. \ref{hbc_basic}. In this particular type of communication two wearable devices placed on two different locations of the human-body communicate among themselves by utilizing the conductivity property of the human body and using it as the communication medium. Capacitive HBC \cite{zimmerman_hbc, maity_bio_phys} is the most widely used form of HBC, where the signal is coupled in a single ended manner in the transmitter end and also received in a similar single ended way at the receiver end. In this scenario, the body provides the forward path of communication between the devices. The return path is formed through the parasitic capacitance between the ground planes of the transmitter and receiver. The overall channel response is strongly dependent on the parasitic return path capacitance. The other primary factor, which determines the channel response, is the termination on the receiver end. A 50$\Omega$ termination at the receiver end results in high loss at low frequencies and hence a high pass response. However, a high impedance capacitive termination at the receiver end enables  a flat-band response with low frequency roll off at frequencies $<$100kHz. For a high impedance termination, the channel loss is dependent on the ratio of the termination capacitance and the return path capacitance. However, for typical values of parameters the channel loss varies from 40-60 dB making it a lossy channel which is flat-band until frequencies as low as 100kHz.

Note that, for the above mentioned two applications or in general any broad-band communication technique where the flat-band or low-frequency channel loss is sufficiently high ($> 20$ dB), the transmitted signal gets highly attenuated while reaching the RX front-end. Hence, voltage mode signaling with rail-to-rail transmitted signal swing should be utilized as opposed to current-mode signaling. Also, due to the low signal swing at the input of the RX, a simple non-return to zero (NRZ) modulation scheme shows superior BER performance over multi-level schemes such as four-level pulse amplitude modulation (PAM-4), duobinary etc. In NRZ communication, the most important component of a RX is the clocked comparator or sampler which distinguishes between the two levels of NRZ data and detects the transmitted bit. In both the aforementioned applications, integrating front-end has been utilized for serving specific purposes associated with the particular type of channel, i.e. to  deal  with self-resonance frequency (SRF)  in  proximity  communication\cite{proximity}  and  cancelling  environmental interference in human-body communication\cite{HBC_CICC}. However, the fact that \textit{current-integrating amplifiers} or \textit{integrators} can be utilized as gain elements with sufficiently lower power consumption, remains relatively unexplored in literature with a dearth of closed-form equations capturing the same. The following analyses of different RX architectures deal with an extensive analysis of different signaling blocks followed by finding the best achievable performances of different architectures on increasing the channel loss. Also, for simplifying the analysis, high-frequency roll-off (if any) of the wireline-like channel has been ignored which alleviates the need of any equalizer circuit in the RX front-end.

\section{Architectural choices}

As mentioned in the previous section, for broad-band NRZ communication, a sampler at the RX front-end samples the received signal at each clock-edge and converts it into full-swing bit-pattern. Hence, a sampler or clocked comparator serves as the simplest RX. However, for practical samplers (e.g. a strongARM latch\cite{SAL1}) the sampling frequency is limited by the input signal swing which decreases with increase in channel-loss. Also, for applications with very high channel-loss, input referred noise of the sampler may become comparable with the signal and can degrade the bit-error rate (BER) drastically. A low-nose amplifier (LNA) can be used before the sampler which serves two important purposes, i.e. it amplifies the RX input signal and exhibits significantly lower input referred noise. However, being a continuous time amplifier its power consumption is large and increases linearly with the required bandwidth. A \textit{current-integrating amplifier} or \textit{integrator}\cite{INT_ISSCC, INT_JSSC_2009, chintan_int_JSSC}, on the other hand can provide gain comparable to an LNA with lower power consumption but at the cost of relatively higher input-referred noise.

Fig. \ref{signaling_blocks} summarizes the typical performance of all these blocks (in 65 nm CMOS) in terms of gain, input referred noise and power consumption for 1 Gbps data rate (i.e. 1 GHz clock frequency). Various RX front-end topology based on these three key signaling blocks are shown in Fig. \ref{archi} with the sampler as a mandatory part in each topology. In the following sections detailed analyses for all these signaling blocks are done and optimum performance of each topology for different channel losses are estimated.

\section{Analysis of signaling blocks}
\subsection{Sampler or clocked comparator}

\begin{figure}[!]
\centering
\includegraphics[height=2.2in]{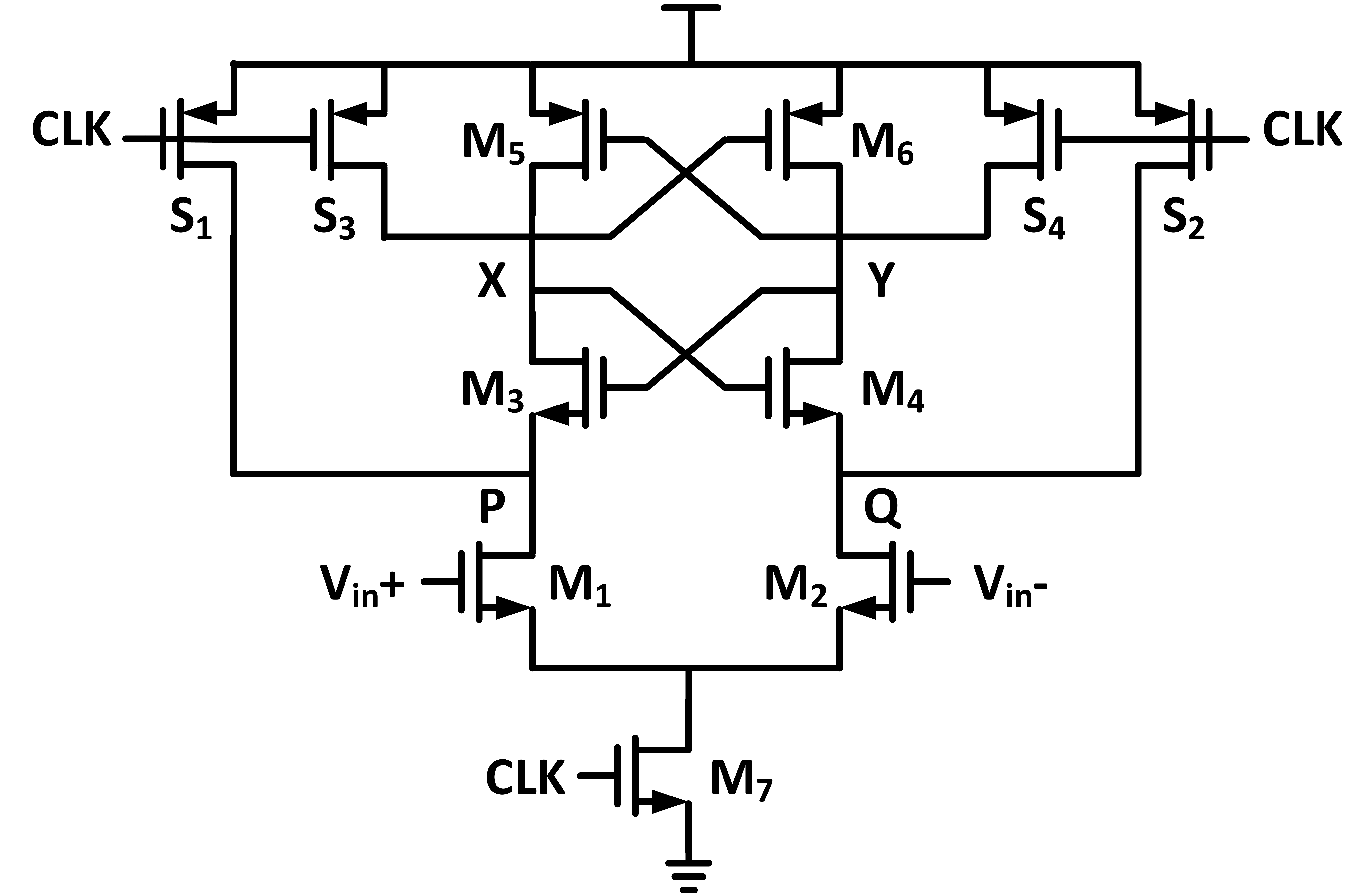}
\caption{Widely used strongARM latch topology\cite{Razavi_SAL}}
\label{SAL_ckt}
\end{figure}

\begin{figure}[!]
\centering
\subfigure[]{
\includegraphics[width=3in]{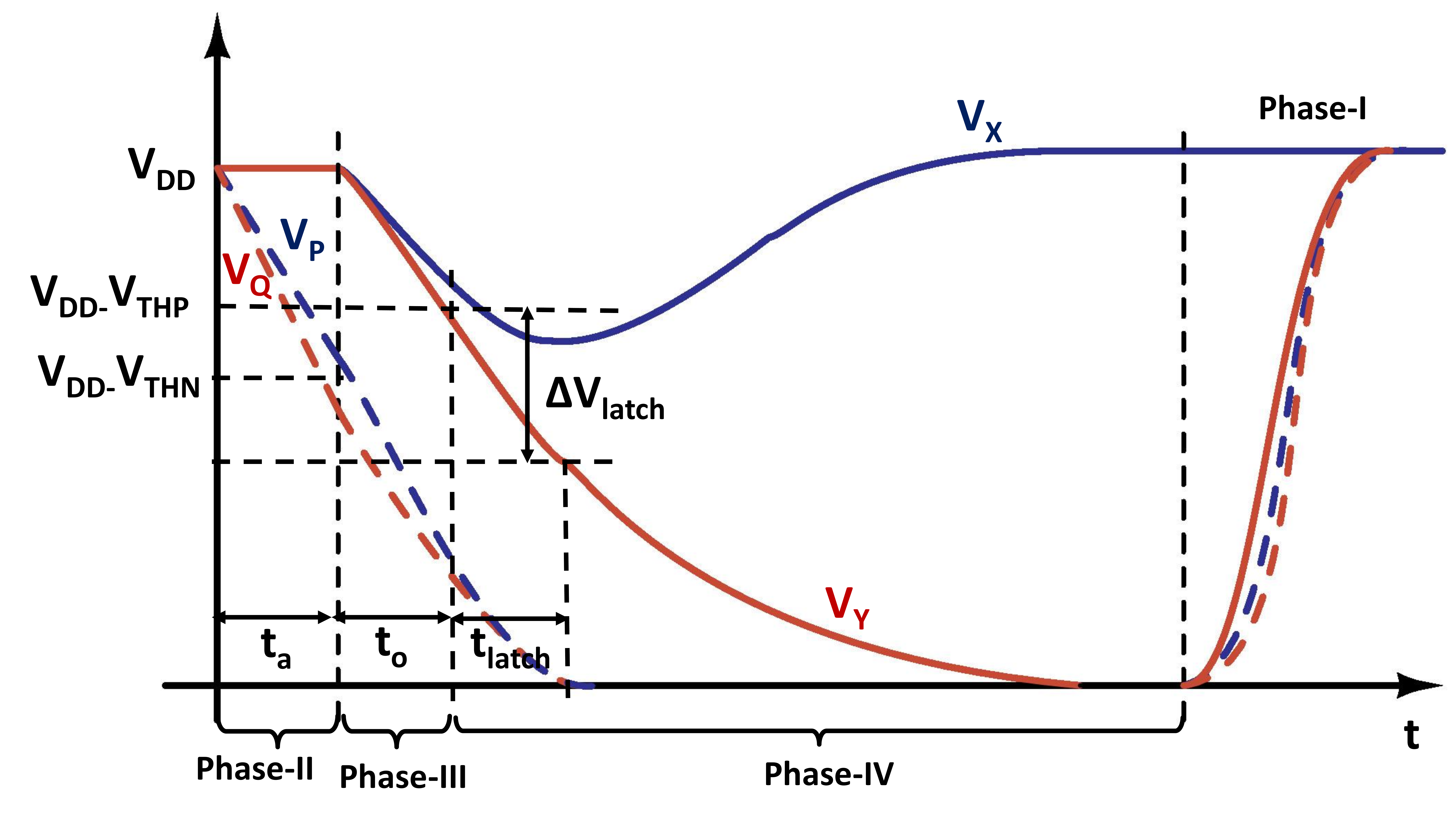}
\label{SAL_trans}
}
\subfigure[]{
\includegraphics[height=2in]{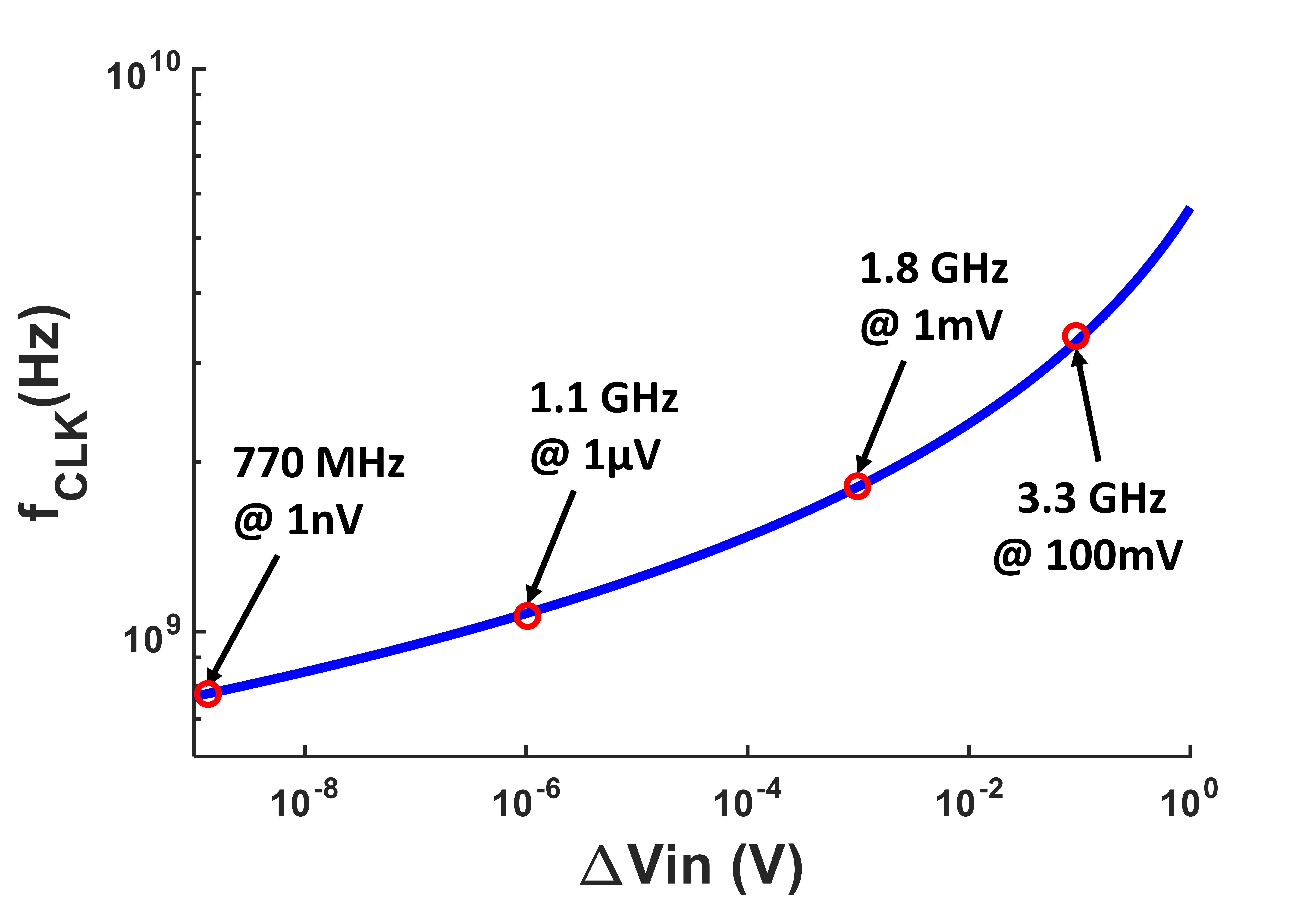}
\label{SAL_freq}
}
\caption{(a) Transient behaviour of a strongARM latch in the sensing phase (i.e. phase-II,III and IV) showing the transition between different phases of operation, (b) Variation of the maximum operation frequency of strongARM latch with input voltage}

\end{figure}

StrongARM latch is the most common type of sampler widely used in different applications including wireline receivers, analog-to-digital converters and memory bit-line detectors. The reason for its widespread popularity is zero static power consumption and rail-to-rail output swing.

The strongARM latch shown in Fig. \ref{SAL_ckt} has four phases of operation\cite{Razavi_SAL}. In the reset phase (phase-I), $CLK$ is low and nodes $P,Q, X$ and $Y$ are pre-charged to $V_{DD}$.  When $CLK$ goes high, amplification phase begins and the input differential voltage at the inputs of $M_1$ and $M_2$ gets converted to differential drain current which is integrated at the parasitic capacitances at nodes $P$ and $Q$ amplifying the input signal until $V_P$ and $V_Q$ drop to $V_{DD}-V_{THN}$ (Fig. \ref{SAL_trans}). At this point $M_3$ and $M_4$ are turned on, output nodes $X,Y$ start discharging and the circuit enters into the third phase with continuing amplification by $M_1$ and $M_2$ and a little regenerative gain provided by $M_3$ and $M_4$. The final regeneration phase begins when nodes $X$ and $Y$ drop below $V_{DD}-V_{THP}$ turning $M_5$ and $M_6$ on. To understand the timing performance of the strongARM latch for applications with high channel loss where input differential voltage can be quite small, \textit{one must carefully consider the dependencies of duration of each phase over the input voltage and also the total input referred noise of the strongARM latch}.
\subsubsection{Transient performance (Latching time consideration)}

Fig. \ref{SAL_trans} shows the transient behaviour of the strongARM latch in the sensing phase (i.e. when $CLK$ becomes high). From the analyses in \cite{latch_eq} and \cite{Razavi_SAL_tut}, duration of different phases can be expressed as

\begin{equation}
    t_a=\frac{2C_{P,Q}V_{TH3,4}}{I_O}
    \label{eq_ta}
\end{equation}

\begin{equation}
    t_o=\frac{2C_{X,Y}V_{TH5,6}}{I_O}
\end{equation}

\begin{equation}
    t_{latch}=\frac{C_{X,Y}}{g_{m,latch}}\ln\mleft(\frac{1}{V_{TH5,6}}\sqrt{\frac{I_O}{2\beta}}\frac{\Delta V_{latch}}{\Delta V_{IN}}\mright)
    \label{eq_tlatch}
\end{equation}

where $I_O(=g_{m1,2}V_{ov1,2}/2)$ is the quiescent current provided by $M_7$ once $CLK$ goes high, $g_{m,latch}$ is the sum of the transconductances of $M_3$ and $M_5$ in the regeneration or latching phase (i.e. phase-III), $\beta$ is the transconductance parameter of $M_1$ and $M_2$ and $\Delta V_{IN}$ is the input differential voltage to the strongARM latch. Note that the regeneration phase is characterized by $t_{latch}$ which in turn is governed by $\Delta V_{latch}$ as shown in \cite{latch_eq}. As the output of the strongARM latch needs to be sampled by a D-flip flop before the reset phase starts, sufficient time should be provided in the regeneration phase so that the differential outputs can reach $V_{DD}$ and $GND$ respectively. Considering this fact, duration of phase-(II+III+IV) is conservatively chosen to be $3\times(t_a+t_o+t_{latch})$ which gives the minimum time period of $CLK$ to be
\begin{equation}
    T_{CLK,min}=\frac{1}{f_{CLK,max}}=6(t_a+t_o+t_{latch})
\end{equation}

From eq. (\ref{eq_tlatch}), it can be seen that as the input differential voltage $\Delta V_{IN}$ reduces, the maximum operating frequency of the strongARM latch decreases logarithmically. Fig. \ref{SAL_freq} shows the variation of maximum operating frequency ($f_{CLK,max}$) with $\Delta V_{IN}$ obtained by extracting all the parameters in eq. ($1$)-($3$) for a typical design in 65 nm CMOS technology. As can be seen, for very small $\Delta V_{IN}$ ($\approx 1$ nV), $f_{CLK,max}$ can be as small as $0.8$ GHz and for larger $\Delta V_{IN}$ ($\approx 100$ mV), the value reaches up to $3.3$ GHz. From Fig. \ref{SAL_freq}, it may seem that with signal amplitude of even a few $\mu$V, the strongARM latch can be operated at a speed close to $1$ GHz, but practically for sub-mV  signal swing, final decision will be significantly affected by the internal noise of strongARM latch. Hence, it is important to find out the total input referred noise which is addressed in the following subsection.

\subsubsection{Noise performance (SNR consideration)}
It is interesting to note that in the reset phase, the pre-charge action of the switches ($S_1-S_4$) nullifies effect of all the noise contributed by different transistors. It is when the $CLK$ goes high, that the noise contributions of different transistors come into picture. Moreover, most of the input referred noise originates from $M_1$ and $M_2$ in the amplification phase\cite{Razavi_SAL_tut} because all other transistors start acting after phase-II when a significant gain has already accrued between nodes $P$ and $Q$ which then gets regenerated in rest of the phases. Hence, the input referred noise of strongARM latch would be simply the output referred noise at the end of amplification phase divided by the gain of the amplification phase given by $g_{m1,2}t_a/C_{P,Q}$. In \cite{SAL_noise}, a stochastic analysis of this noise has been done. Here we show a time domain analysis based on the \textit{ergodicity property} of thermal noise.

Note that in the amplification phase, as $M_3$ and $M_4$ are turned off, strongARM latch behaves like an integrator, integrating the drain currents of $M_1$ and $M_2$ over the parasitic capacitance $C_P$ and $C_Q$ at nodes $P$ and $Q$ respectively. Hence, assuming the output resistance at nodes $P,Q$ to be very large, the final differential output referred noise at the end of the amplification phase can be found by integrating the differential channel noise current $i_n$ ( whose $PSD=8KT\gamma g_{m1,2}$) of $M_1$ and $M_2$ for a duration of $t_a$. This gives the final noise voltage to be
\begin{equation}
    V_{n,O}=\frac{1}{C_{P,Q}}\int_{0}^{t_a}i_n(t)dt
    \label{SAL_noise_int}
\end{equation}

To evaluate this integral, let us first assume $i_n(t)$ to be a $sine$ wave with amplitude $A$, frequency $f$ and initial phase $\phi$, i.e. $i_n(t)=Asin(2\pi ft+\phi)$. For this simplest scenario, $V_{n,O}$ can be found out to be
\begin{equation}
    V_{n,O}=\frac{A(cos(\phi)-cos(2\pi ft_a+\phi))}{2\pi fC_{P,Q}}=\frac{A}{C_{P,Q}}.TF_\phi(f)
    \label{int_tfeq}
\end{equation}

\begin{figure}[!]
\centering
\includegraphics[height=2.2in]{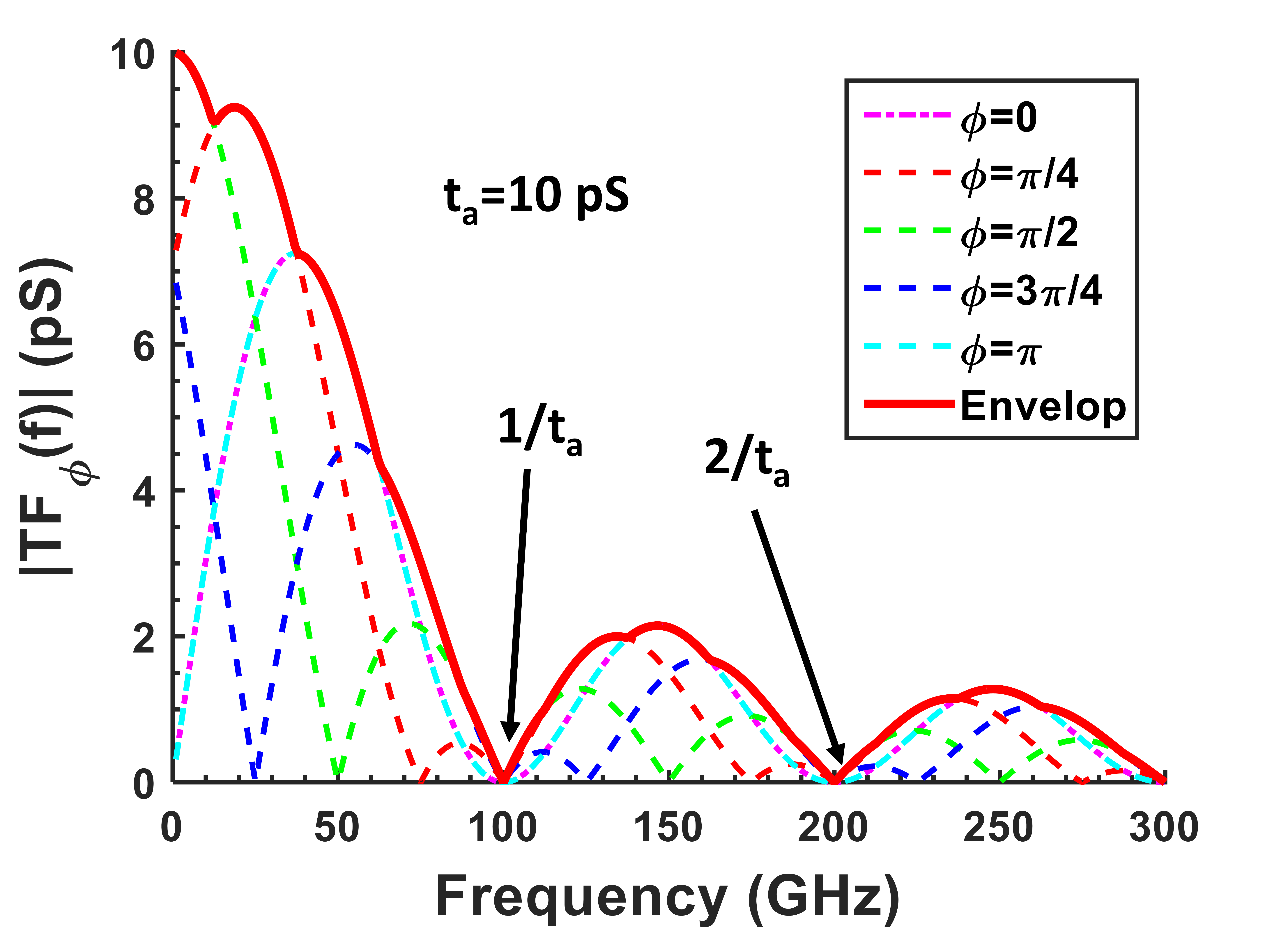}
\caption{Plot of the function $TF_\phi(f)$ which shapes the noise spectrum in strongARM latch. In this case, the envelop is shown considering only $5$ different values of $\phi$ }
\label{int_tf}
\end{figure}

\begin{figure}[!]
\centering
\includegraphics[height=2.2in]{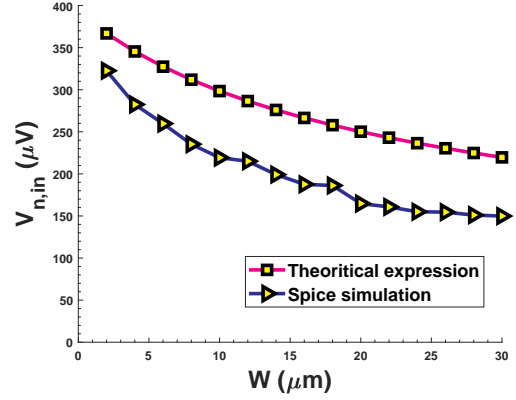}
\caption{Variation of input referred noise of strongARM latch for different widths of $M_1$ and $M_2$. Noise is calculated both from the theory (eq. \ref{i/p_ref_noise}) and from spice simulation following the method in \cite{Razavi_SAL_tut}}
\label{SAL_noise_w}
\end{figure}

From eq. (\ref{int_tfeq}), it can be seen that if $i_n(t)$ be a sinusoid with frequency $f$ and initial phase $\phi$, result of the integral in eq. (\ref{SAL_noise_int}) would be  $TF_\phi(f)$ times its amplitude. But in reality, $i_n(t)$ in the time interval $0$ to $t_a$, contains all the frequency components and hence different frequency components would have different multiplication factor depending on their initial phases. Hence, the final noise voltage can be found by summing the contributions of all the frequencies present in $i_n(t)$. Fig. \ref{int_tf} shows the plot of $TF_\phi(f)$ for $5$ different values of $\phi$. The rms value of the component of $i_n(t)$ obtained by passing it through a band-pass filter of bandwidth $\Delta f$ centered around frequency $f$ can be given by $\sqrt{8KT\gamma g_{m1,2}\Delta f}$. But as the initial phase corresponding to this component at frequency $f$ can eventually be anything, by pessimistic assumption we can consider the multiplication factor to be $max(TF_\phi(f):\phi \epsilon [0:2\pi])$ which is essentially the envelop ($TF_{env}(f)$) of all the curves governed by different $\phi$. Hence, considering the noise contribution of all the components in $i_n$, the final rms noise voltage square ($V^2_{n,O}$) can be expressed as

\begin{equation}\label{o/p_ref_noise_square}
\begin{split}
      V^2_{n,O} & =\sum_{f}\mleft(\frac{\sqrt{8KT\gamma g_{m1,2}\Delta f}}{C_{P,Q}}\times TF_{env}(f)\mright)^2\\
      & =\frac{8KT\gamma g_{m1,2}}{C^2_{P,Q}}\times\int_{0}^{\infty}TF^2_{env}(f)df
\end{split}
\end{equation}

Hence, the input referred noise can be given by
\begin{equation}\label{i/p_ref_noise_square}
\begin{split}
      V^2_{n,in} & =V^2_{n,in}/\mleft(\frac{g_{m1,2}t_a}{C_{P,Q}}\mright)^2\\
      & =\frac{8KT\gamma}{g_{m1,2}t^2_a}\times\int_{0}^{\infty}TF^2_{env}(f)df
\end{split}
\end{equation}
The integral in eq. (\ref{i/p_ref_noise_square}) can be numerically evaluated to be $t_a/2$ which gives the final input referred noise expression as
\begin{equation}\label{i/p_ref_noise}
    V^2_{n,in}=\frac{4KT\gamma}{g_{m1,2}t_a}
\end{equation}
which exactly matches with the stochastic analysis result in \cite{SAL_noise}. Moreover, on replacing $t_a$ with ($\ref{eq_ta}$) and substituting the value of $I_O$ one gets the expression for input referred noise as
\begin{equation}\label{i/p_ref_noise}
    V^2_{n,in}=M \frac{KT}{C_{P,Q}}
\end{equation}
where $M=\gamma V_{TH3,4}/V_{ov1,2}$. Note that, the noise term has an usual $KT/C$-form with an additional factor-$M$. To validate this theory, input referred noise of strongARM latch has been obtained in spice simulation following the method described in \cite{Razavi_SAL_tut}. Fig. \ref{SAL_noise_w} compares the spice result with the theoretical expression plotted by extracting transistor parameters in 65 nm CMOS.

\subsection{Low noise amplifier}

\begin{figure}[!]
\centering
\includegraphics[height=1.5in]{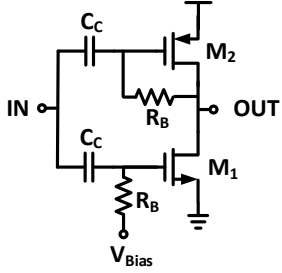}
\caption{Circuit diagram of low noise amplifier with self-biased load.}
\label{LNA_ckt}
\end{figure}

\begin{figure*}[!]
\centering
\subfigure[]{
\includegraphics[height=1.8in]{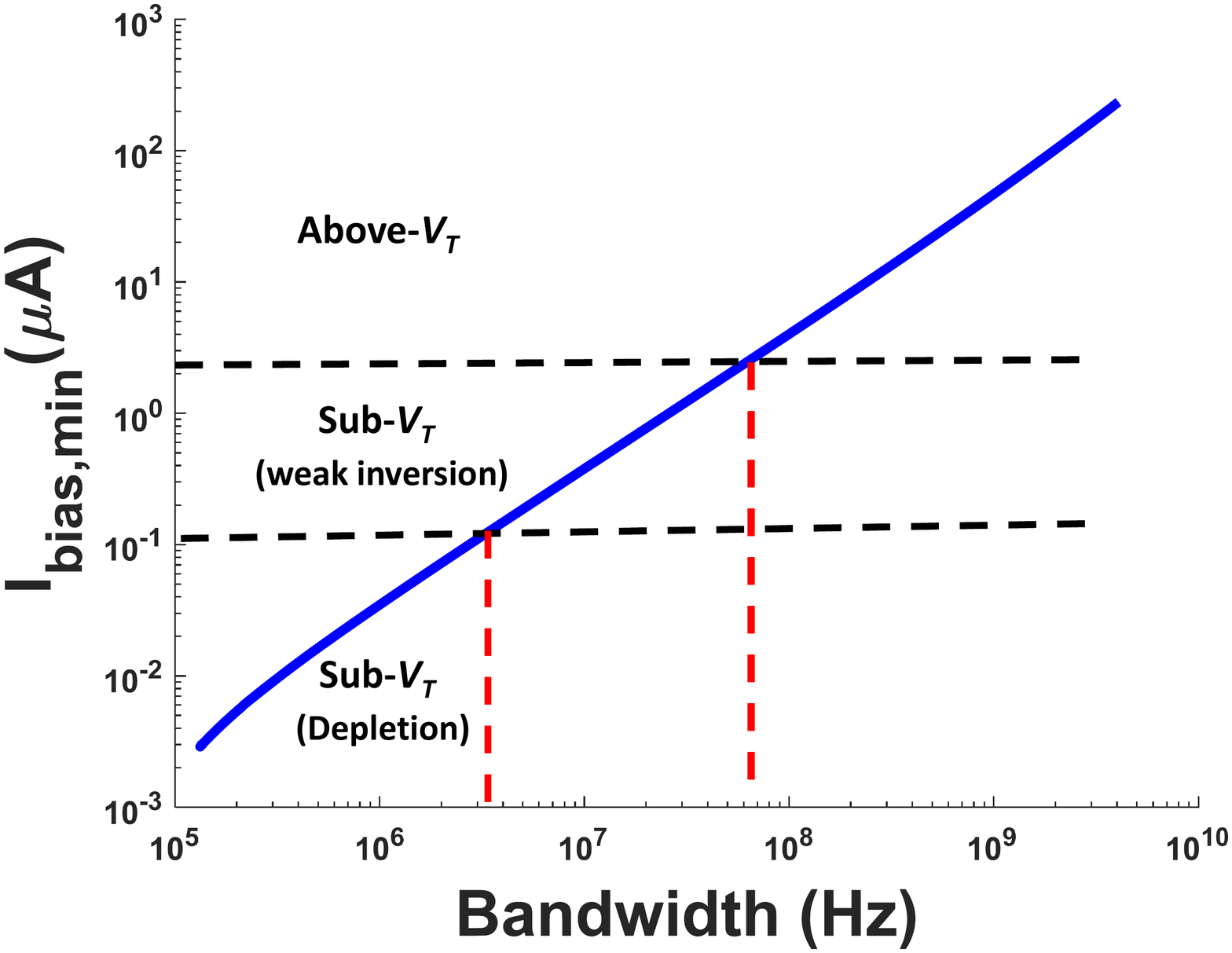}
\label{curr_LNA}
}
\hspace*{-0.3in}
\subfigure[]{
\includegraphics[height=1.8in]{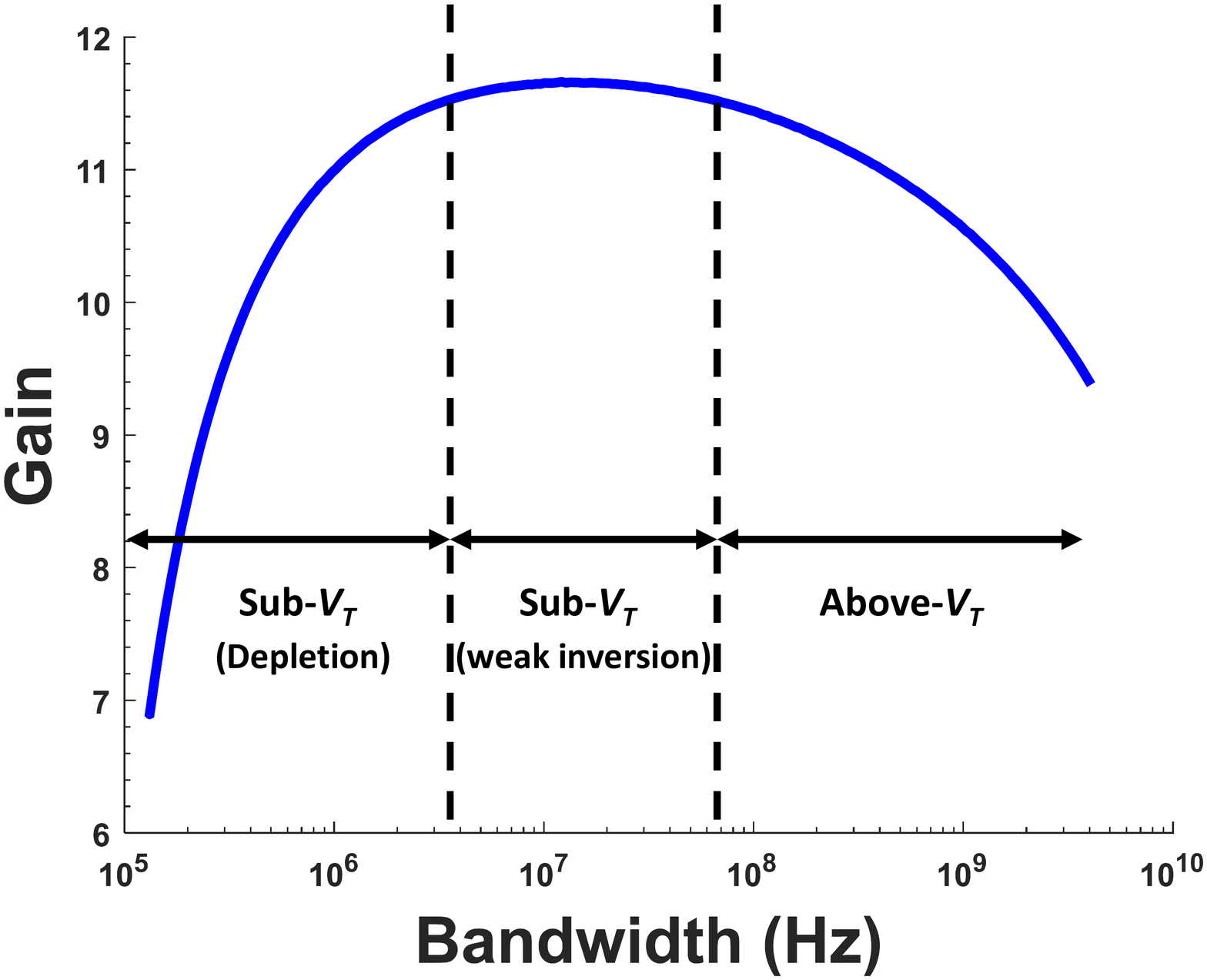}
\label{gain_LNA}
}
\hspace*{-0.3in}
\subfigure[]{
\includegraphics[height=1.8in]{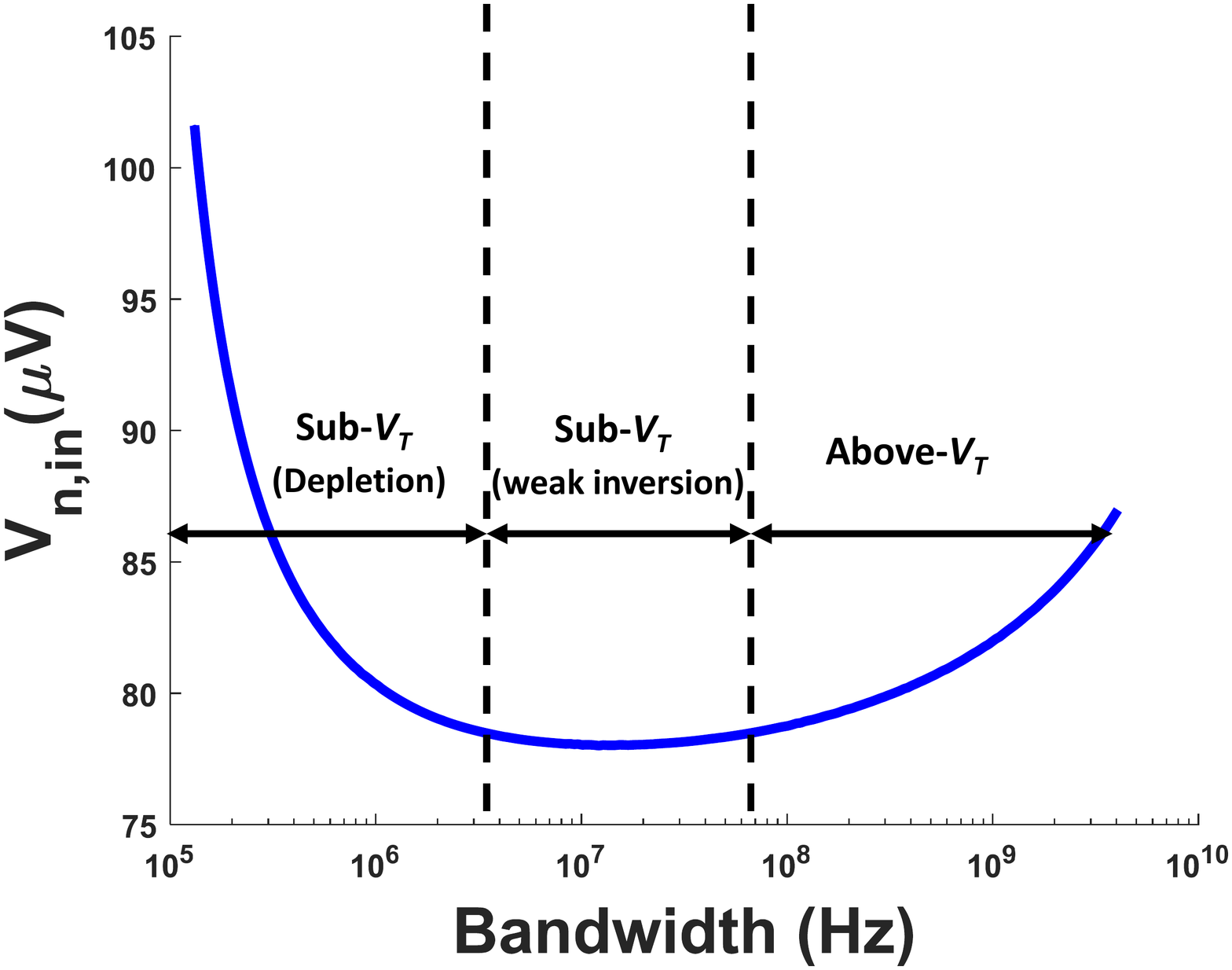}
\label{noise_LNA}
}
\caption{(a) Minimum bias current required for the LNA in Fig. \ref{LNA_ckt} for different target bandwidths (b) gain of the LNA for different target bandwidths with minimum bias current (c) input referred noise of the LNA for different target bandwidths with minimum bias current.}
\label{LNA_plots}
\end{figure*}

As mentioned in section.III, a low noise amplifier serves two important purposes: i) it provides a gain to the inbound signal and ii) offers much lower input referred noise. The most commonly used broadband topology of a low-noise amplifier (LNA) is shown in Fig. \ref{LNA_ckt} where $R_B$ is a large resistor (realized by off-transistor in this case). The mid-band gain of this LNA can be given by 
\begin{equation}
    A_{LNA}=(g_{m1}+g_{m2})(r_{o1}||r_{o2})
\end{equation}
where $g_{m1,2}$ is the transconductance of $M_{1,2}$ and $r_{o1,2}$ is its drain to source resistance. Hence the gain of the LNA is solely determined by the intrinsic gains of transistors $M_1$ and $M_2$. Also, the input referred noise of this LNA can be expressed as
\begin{equation}
    V_{n,in,LNA}=\sqrt{\frac{4KT\gamma}{g_{m1}+g_{m2}}\times B}
    \label{LNA_noise_eq}
\end{equation}

where $B$ is the bandwidth of the LNA which depends on the bias current ($I_{bias}$) and effective load capacitance. Note that the bandwidth requirement of the LNA comes from the input data-rate, i.e. the bandwidth of the LNA should be larger than or equal to the data rate to avoid any signal distortion causing inter-symbol interference (ISI). For the analysis in Fig. \ref{LNA_plots} a typical design setup in $65$ nm CMOS is used where sizes of the transistors are kept constant (width of $M_1$ is $24$ $\mu$m and that of $M_2$ is $48$ $\mu$m) and an external load capacitance ($C_L$) of $10$ fF has been chosen. Fig. \ref{curr_LNA} shows the minimum $I_{bias}$ required for the LNA as a function of its bandwidth. It is important to note that as the bandwidth requirement of the LNA goes down, operating region of the transistors moves from above-$V_T$ to weak inversion and finally to depletion region. And hence, the power consumption goes down more-or-less linearly with reduction in target bandwidth. Fig. \ref{gain_LNA} shows the variation of the mid-band gain of the LNA with bandwidth, assuming the minimum bias current in the LNA for each target bandwidth. Note that for target bandwidths where transistors are in above-$V_T$ region of operation, gain increases with reduction in bias current as $g_m$ is proportional to $\sqrt{I_{bias}}$ and $r_o$ is inversely proportional to $I_{bias}$ in this region. When the transistors go to week inversion, $g_m$ becomes proportional to $I_{bias}$ and hence, the gain remains almost constant. Finally, in the depletion region gain falls with reduction in current as $r_o$ becomes comparable to $R_B$. On the other hand, the input referred noise (in Fig. \ref{noise_LNA}) behaves exactly the opposite to the gain, for different target bandwidths, as expected from the noise expression in eq. (\ref{LNA_noise_eq}). On comparing the noise performance of strongARM latch in Fig. \ref{SAL_noise_w} to that of the LNA in Fig. \ref{noise_LNA}
shows significantly lower input referred noise for the LNA. This, together with the gain plot in Fig. \ref{gain_LNA} substantiate the use of the LNA before the sampler for achieving higher data rate in applications with high channel loss.

\subsection{Integrating amplifier or integrator}

\begin{figure}[!t]
\centering
\includegraphics[height=4in]{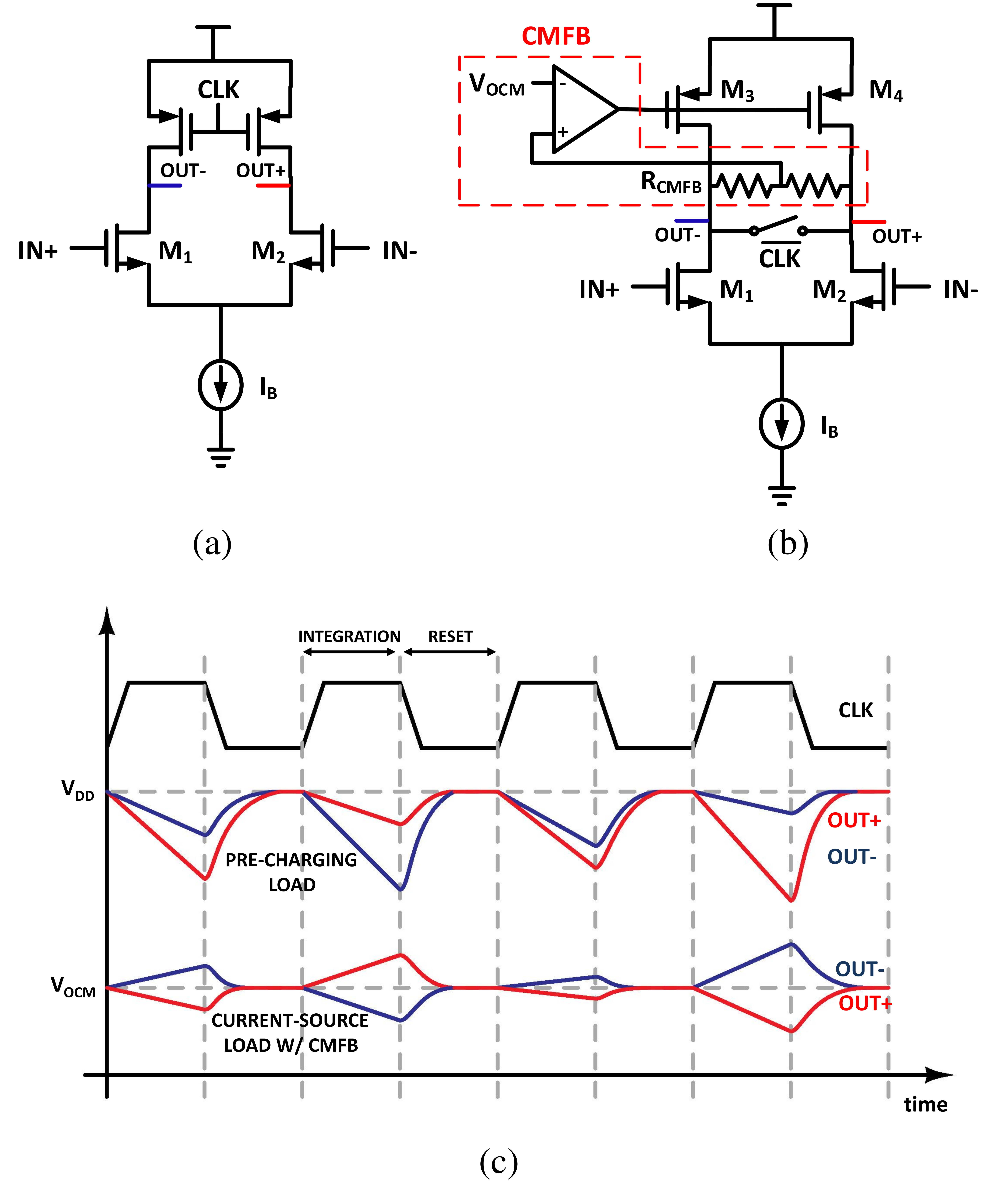}
\caption{(a) Integrator based on pre-charging load \cite{int_ISSCC}, (b) modified integrator based on current-source load with common-mode feedback \cite{chintan_int_JSSC}, (c) comparison of output waveform for both the integrators. Output common-mode voltage of the one with pre-charging loads keeps on decreasing, whereas the other one has a constant common-mode voltage. }
\label{INT_ckt}
\end{figure}

An integrating amplifier can be used in the RX front-end to provide gain to the received signal before sampling with significantly lower power consumption than an LNA. Fig. \ref{INT_ckt}(a) shows the circuit diagram of the integrator which utilizes pre-charging loads \cite{int_ISSCC}. When $CLK$ is low, the PMOS switches are on, which pre-charge the output nodes to $V_{DD}$. As $CLK$ goes high, the output nodes start discharging and depending on the input voltage difference, a finite voltage difference is created between the output nodes which is then sampled by the sampler at the end of integration period. Assuming large output resistance, the ratio of the output and input voltage difference or the voltage gain of the integrator can be expressed as
\begin{equation}
    A_{int}=\frac{g_{m1,2}T_{int}}{C_L}
    \label{gain_INT}
\end{equation}
where $g_{m1,2}$ is the transconductance of $M_{1,2}$, $T_{int}$ is the period of integration which equals to half of the clock period and $C_L$ is the equivalent load capacitor at the output nodes. Note that the integrator senses the input data only in the integration phase and hence, for clock frequency same as the data-rate, the integrator integrates the input data for only half the bit-period. However, an half-rate architecture \cite{int_ISSCC} can be used where two parallel integrators work on complementary clock signals with clock frequency as half the data-rate. 

\begin{figure}[!t]
\centering
\includegraphics[height=2in]{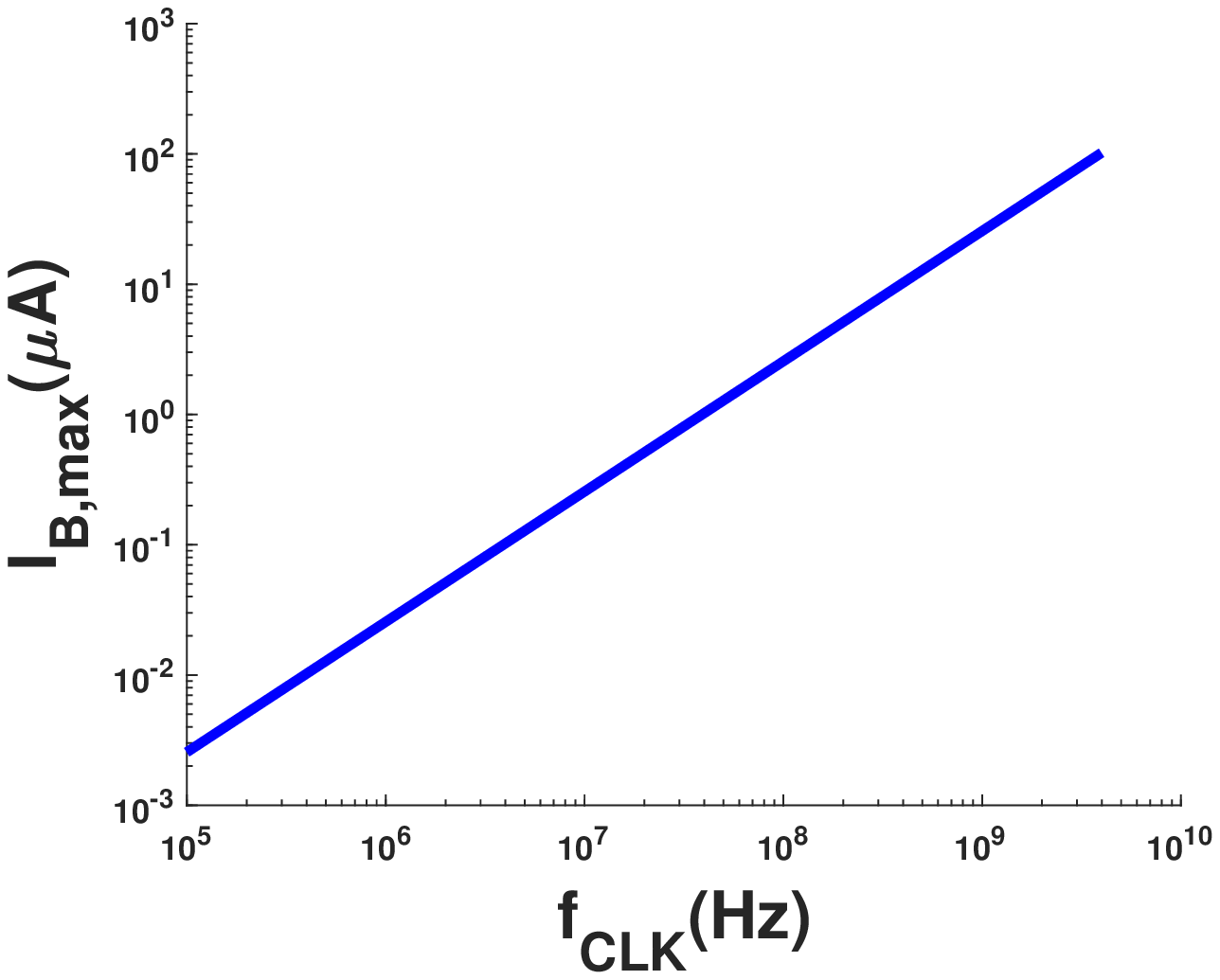}
\caption{Variation of the maximum allowable current of the integrator in Fig. \ref{INT_ckt} with clock frequency.}
\label{curr_INT}
\end{figure}

It is important to observe that as the integrator output is fed to the sampler realized by a strongARM latch, the common mode voltage of the output nodes at the end of integration phase should be high enough for proper operation of the strongARM latch. \cite{latch_eq} shows that $0.7 V_{DD}$ is the optimum input common mode voltage for the strongARM latch in terms of speed and yield. However, for 65 nm CMOS, we found that the input common mode voltage can go down to $\sim 0.6 V_{DD}$ without degrading the speed and yield significantly. With this observation, the maximum bias current ($I_{B,max}$) in the integrator for $V_{DD}=1$V can be expressed as
\begin{equation}
    I_{B,max}=\frac{(0.4V_{DD})(2C_L)}{T_{int}}=\frac{0.8C_L}{T_{int}}
    \label{curr_max_int}
\end{equation}
Fig. \ref{curr_INT} shows the variation of $I_{B,max}$ with clock frequency for $C_L=4$ fF, which gives $I_{B,max}\approx 20\mu$A for a $1$ GHz clock frequency. The common-mode droop problem associated with the integrator in Fig. \ref{INT_ckt}(a) is handled in \cite{int_CM_droop1} by a separate common-mode boosting circuit using capacitive coupling, whereas \cite{int_CM_droop2} addresses the same by adding a common-mode current during the integration phase. In \cite{chintan_int_JSSC} the output common-mode voltage is kept constant by using current-source loads and a common-mode feedback (CMFB) circuit as shown in Fig. \ref{INT_ckt}(b). The corresponding output waveform is shown in Fig. \ref{INT_ckt}(c).

\begin{figure}[!]
\centering
\includegraphics[height=1.2in]{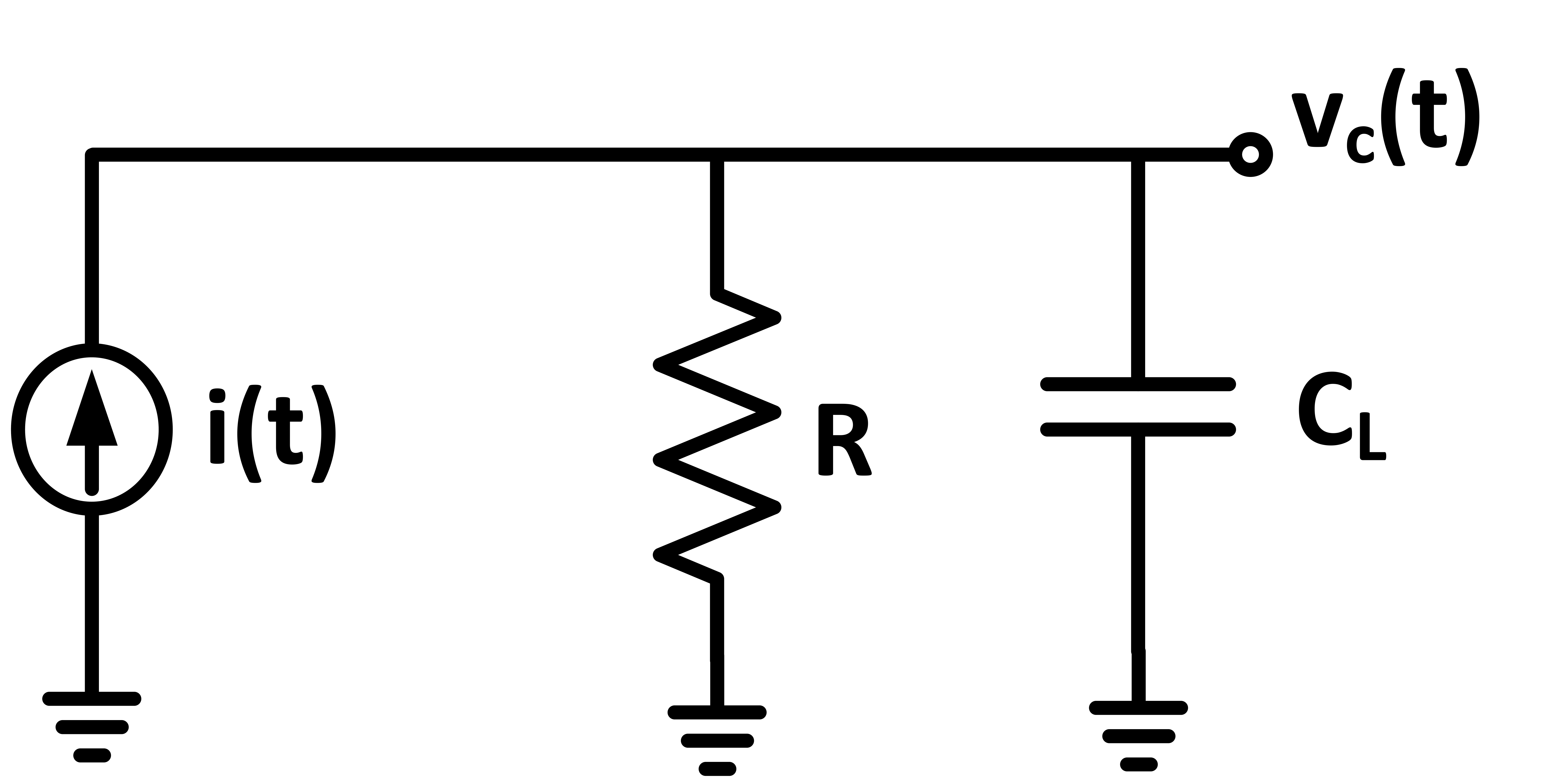}
\caption{Equivalent circuit of the integrator (shown in Fig. \ref{INT_ckt} (a,b)) in the integration phase}
\label{int_model}
\end{figure}

From the gain expression in eq. (\ref{gain_INT}) it might seem that the gain of the integrator in Fig. \ref{INT_ckt}(a) can go very large for lower clock frequency as $g_{m1,2}\propto \sqrt{I_{B,max}}$ and $T_{int}\propto 1/I_{B,max}$. But practically, the gain will be different from that given by eq. (\ref{gain_INT}) due to the presence of finite output resistance of the integrator. In the following portion, we derive an accurate expression for the gain of the integrator considering the drain to source resistance of $M_{1,2}$ in the integration phase.

In the integration phase, the circuit in Fig. \ref{INT_ckt}(a) can be simplified as shown in Fig. \ref{int_model}, where $i(t)=g_{m1,2}v_{in}$, $R$ is the drain to source resistance of $M_{1,2}$ and $C_L$ is the load capacitance at the output nodes. Now, from Kirchoff's current law, we can write
\begin{equation}
    i(t)=\frac{v_c(t)}{R}+C_L\frac{dv_c(t)}{dt}
\end{equation}
on solving this differential equation and applying the initial pre-charge condition $v_c(0)=0$, we can write
\begin{equation}
    v_c(t)=\frac{e^{t/RC_L}}{C_L}\int_0^t e^{\tau /RC_L}i(\tau)d\tau
    \label{int_out_eq}
\end{equation}
finally substituting $i(\tau)$ with $g_{m1,2}v_{in}$ we can express the gain as
\begin{equation}
    A_{int}=\frac{v_c(T_{int})}{v_{in}}=g_{m1,2}R\mleft(1-e^{-T_{int}/RC_L}\mright)
    \label{gain_INT_real}
\end{equation}

\begin{figure}[!]
\centering
\includegraphics[height=2in]{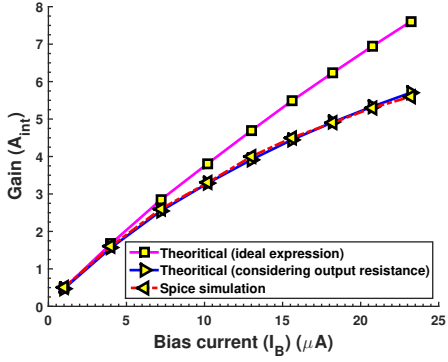}
\caption{Variation of integrator gain with bias current for $f_{CLK}=1$ GHz. Note that, the plot obtained from the modified expression in eq. (\ref{gain_INT_real}) exactly matches with that obtained from spice simulation.}
\label{int_gain_ideal_modified}
\end{figure}

\begin{figure}[!]
\centering
\includegraphics[height=2in]{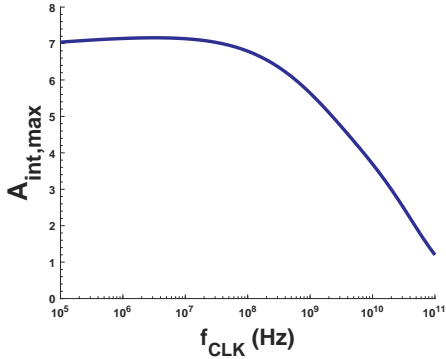}
\caption{Variation of the maximum gain of the integrator with clock frequency. As the maximum gain is mainly governed by the intrinsic gain of the transistors (eq. (\ref{gain_INT_real_max})), it saturates for lower clock frequencies as the transistors enter into sub-threshold regime.}
\label{gain_max_int}
\end{figure}

\begin{figure}[!]
\centering
\includegraphics[height=2in]{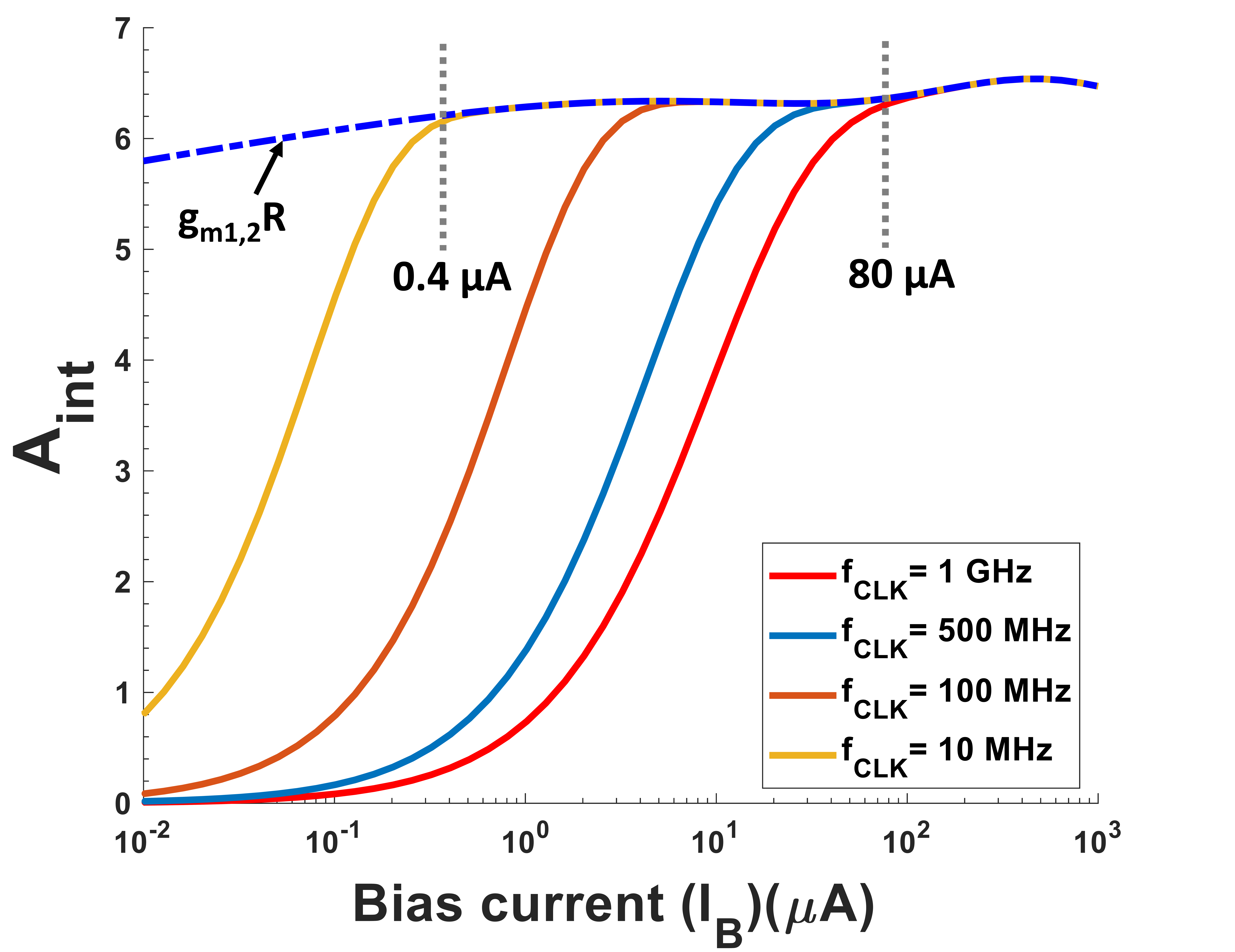}
\caption{Variation of gain of the current-source load based integrator for different clock frequencies}
\label{gain_cs_int}
\end{figure}

Note that if $T_{int}<<RC_L$, expression in eq. (\ref{gain_INT_real}) matches to the ideal gain expression in eq. (\ref{gain_INT}). Fig. \ref{int_gain_ideal_modified} compares the integrator gain obtained from the ideal (eq. (\ref{gain_INT})) and modified (eq. (\ref{gain_INT_real})) expressions to that obtained from spice simulation, for a clock frequency ($f_{CLK}$) of $1$ GHz. It can be seen, there is an excellent match between the modified theoretical expression and spice simulation. Note from Fig. \ref{int_gain_ideal_modified} that, as the bias current $I_B$ reduces for a fixed $T_{int}$, value of $R$ increases (as $R\propto 1/I_B$) and hence the two gain expressions give same results for very low $I_B$. Now, for a particular $f_{CLK}$, as the gain increases with $I_B$, the maximum gain can be achieved with $I_B=I_{B,max}$ given by eq. (\ref{curr_max_int}). Also, considering the fact that $R=\frac{2}{\lambda I_B}$ ( $\lambda$ being the channel length modulation parameter) for $I_B=I_{B,max}$ eq. (\ref{gain_INT_real}) becomes
\begin{equation}
    A_{int,max}=g_{m1,2}R\mleft(1-e^{-0.4\lambda}\mright)
    \label{gain_INT_real_max}
\end{equation}

Hence, it can be concluded that the maximum gain of the integrator in Fig. \ref{INT_ckt}(a) for a particular clock frequency ($f_{CLK}$), is mostly governed by the intrinsic gain of the transistors $M_{1,2}$ and it can not be made very high for lower clock frequencies. Fig. \ref{gain_max_int} shows the variation of the maximum gain ($A_{int,max}$) with clock frequency. Note that, the difference in gains of the integrators with pre-charging load (Fig. \ref{INT_ckt}(a)) and current-source load (Fig. \ref{INT_ckt}(b)) arises from the difference in their output resistances ($R$). For the same bias current, $I_B$, the output resistance of the later one is half of the output resistance of the former assuming the same channel-length modulation parameter ($\lambda$) for both PMOS and NMOS devices. This results in a lower gain for the current-source load based integrator compared to the former for the same bias current. Also, as the output common-mode voltage of the current-source load based integrator is constant, the bias current doesn't impose any constraint over the maximum achievable gain. However, note from eq. (\ref{gain_INT_real}) that the maximum achievable gain for the current-source based integrator for a fixed $T_{int}$ (and hence fixed $f_{CLK}$) can not exceed $g_{m1,2}R$ which is half of the intrinsic gain of $M_{1,2}$. Fig. \ref{gain_cs_int} shows the variation of the gain of the current-source load based integrator with bias current, $I_B$ for different clock frequencies ($f_{CLK}$). As it can be seen, the gain increases with increase in $I_B$ and finally converges with $g_{m1,2}R$. Also, the minimum $I_B$ required to converge with $g_{m1,2}R$ decreases with decrease in $f_{CLK}$.

Coming to the noise performance of the integrator, note that the theory of input referred noise described in section IV.A.2 directly applies to the integrator in Fig. \ref{INT_ckt}(a) with $t_a$ being replaced by $T_{int}$. Hence, the input referred noise of the integrator with pre-charging load can be given as
\begin{equation}
    V_{n,in}=\sqrt{\frac{4KT\gamma}{g_{m1,2}T_{int}}}
\end{equation}

However, for the current-source load based integrator, the PMOS current sources $M_3$ and $M_4$ will also contribute to the input referred noise and it can be shown that the overall input referred noise in this case can be expressed as
\begin{equation}
    V_{n,in}=\sqrt{\frac{4KT\gamma}{g_{m1,2}T_{int}}\mleft(1+\frac{g_{m3,4}}{g_{m1,2}}\mright)}
\end{equation}

\begin{figure}[!t]
\centering
\includegraphics[height=2in]{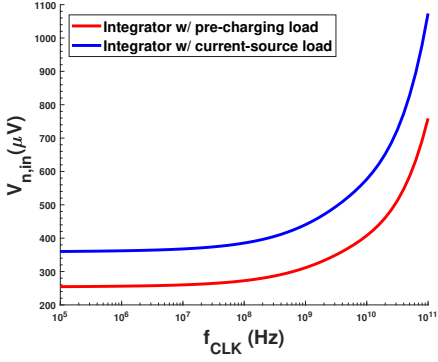}
\caption{Variation of the input referred noise of the integrator with clock frequency}
\label{noise_min_int}
\end{figure}

Fig. \ref{noise_min_int} shows the variation of $V_{n,in}$ with clock frequency for both type of integrtors. Finally, it can be concluded that performance of the pre-charging load based integrator is superior to that of the current-source based integrator in terms of gain, noise performance and power consumption. However, applications where the data rate (and hence the clock frequency) varies widely making a stable output common-mode voltage of the integrator an absolute necessity, the current-source based integrator turns out to be more effective. In all other practical applications, the pre-charging load based integrator gives superior performance. In this work, the pre-charging load based integrator has been used to analyze the performance of all the RX architectures.

\subsection{Multi-integrator cascade: Cascading multiple integrators}
\begin{figure}[!b]
\centering
\includegraphics[height=0.8in]{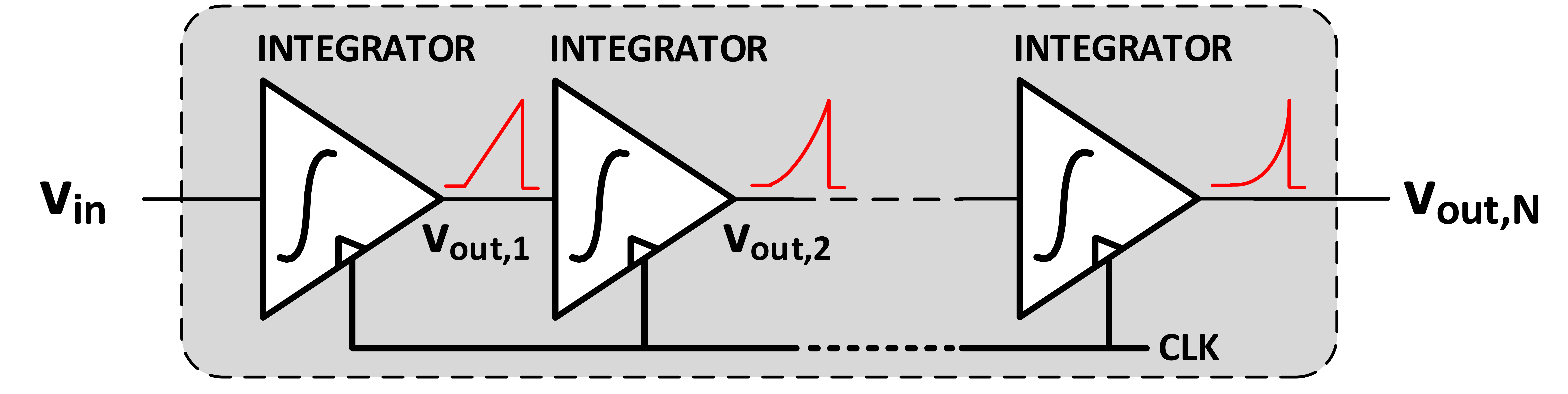}
\caption{Cascading multiple integrators to improve the gain}
\label{multi_int}
\end{figure}

As already seen, a single integrator can typically provide a gain ranging from $4.5-7$ for all frequencies of operation. It will be interesting to see whether the gain can be further enhanced by cascading multiple integrators operating with the same clock signal as shown in Fig. \ref{multi_int}. To understand the behavior of the cascaded integrators in Fig. \ref{multi_int}, let us for the time being ignore the effect of the output resistance. Hence output of the first integrator and the gain can be given by

\begin{figure}[!t]
\centering
\includegraphics[height=2in]{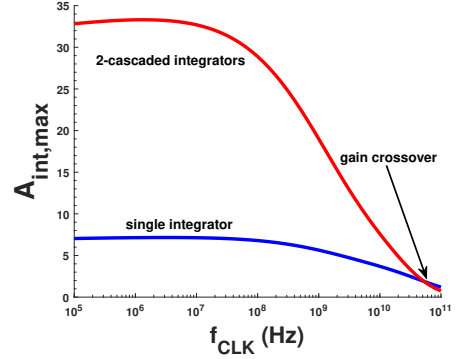}
\caption{Gain improvement with 2 cascaded integrators over a single integrator}
\label{gain_comp_multi_int}
\end{figure}

\begin{equation}
    v_{out,1}(t)=\int_0^t K_iv_{in}dt=K_iv_{in}t
\end{equation}
\begin{equation}
    A_{int,1}=K_iT_{int}
\end{equation}
where $K_i=g_{m1,2}/C_L$. Similarly, output of the second stage and the combined gain of $2$ cascaded integrators can be expressed as
\begin{equation}
     v_{out,2}(t)=\int_0^t K_i(K_iv_{in}t)dt=\frac{K^2_iv_{in}t^2}{2}
\end{equation}
\begin{equation}
     A_{int,2}=\frac{K^2_iT^2_{int}}{2}=A_{int,1}\times \frac{K_iT_{int}}{2}
     \label{2_cas_int_eq}
\end{equation}

Hence, from eq. (\ref{2_cas_int_eq}) it can be observed that in order that gain of $2$ cascaded integrators ($A_{int,2}$) be larger than that of a single integrator, gain of a single integrator ($=K_iT_{int}$) must be greater than 2. Proceeding in the same way it can be shown that the overall gain of $N$-cascaded integrators would be
\begin{equation}
     A_{int,N}=\frac{K^N_iT^N_{int}}{N!}=A_{int,N-1}\times \frac{K_iT_{int}}{N}
     \label{N_cas_int_eq} 
\end{equation}

This is an important result which shows that for a fixed clock frequency (or, equivalently fixed $T_{int}$) and with single integrator gain $A$, cascading $[A]$ (box of $A$) many integrators results in maximum overall gain and the gain starts falling on cascading integrators further. Now, considering the effect of the output resistance of the integrator, using eq. (\ref{int_out_eq}) the overall gain of $2$ cascaded integrators can be expressed as
\begin{equation}
    A_{int,2}=(g_{m1,2}R)^2\mleft(1-\mleft(1+\frac{T_{int}}{RC_L}\mright)e^{-T_{int}/RC_L}\mright)
    \label{2_cas_int_eq_mod}
\end{equation}
which again matches the expression in eq. (\ref{2_cas_int_eq}) if $T_{int}<<RC_L$. Fig. \ref{gain_comp_multi_int} compares the gain of $2$ cascaded integrators obtained from eq. (\ref{2_cas_int_eq_mod}) to that of a single integrator. Note that 2-cascaded integrators can provide much higher gain than the LNA with even lower power consumption than the LNA.

\section{Performance of different architectures for lossy broadband channels}

Based on the detailed analyses of different signaling blocks in the previous section, we are now in a position to compare the performance of each architecture in Fig. \ref{archi} for different channel loss. In the performance analysis of different topologies, a full rate RX architecture has been assumed where the clock frequency is the same as the data rate. The methodology adopted to find the maximum achievable data-rate of each architecture as a function of the channel-loss is delineated below.

\begin{figure}[!t]
\centering
\includegraphics[height=2in]{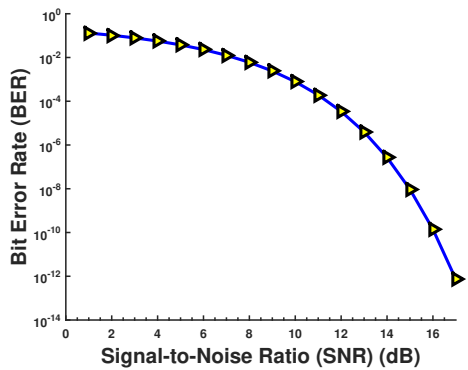}
\caption{Variation of bit-error rate (BER) with signal-to-noise ratio (SNR) for NRZ communication}
\label{BER}
\end{figure}

For any particular topology, as the channel loss ($L$) increases, the input signal swing to the RX ($v_{RX}(L)$) reduces. Let $A_{FE}(f)$ be the gain of the RX front-end which depends on the operating-frequency $f$ (or, equivalently the data-rate). Then, the input voltage of the sampler is given by, $v_{SAL}(f,L)=A_{FE}(f)v_{RX}(L)$. If $g$ be function which maps the input voltage of the strongARM latch ($v_{SAL}$) to its maximum operating frequency (Fig. \ref{SAL_freq}), then a data-rate $f$ is achievable by an architecture \textit{iff} $g(A_{FE}(f)v_{SAL}(L))\geq f$. Hence, the maximum achievable data-rate ($f_{max}$) for an architecture corresponding to a channel loss $L$ is one for which the previous equality holds, i.e. $g(A_{FE}(f_{max})v_{SAL}(L))=f_{max}$. Note that, in this methodology, noise of the front-end has not been considered. However, with reduction in the input signal swing of RX, the input signal-to-noise ratio (SNR) reduces which in turns degrade the bit-error rate (BER) of the final received data. Hence, the total input referred noise of a topology limits the maximum channel loss it can support ($L_{max}$) for a target BER. Fig. \ref{BER} shows the BER vs SNR plot for NRZ communication. For wireline applications, a target BER of $10^{-12}$ is generally used and for mm-wave (or in general wireless) applications the preferred target BER is $10^{-3}$ considering the large loss of wireless channel. However, for wireline-like channels any target BER in this range can be chosen depending on the application and the value of channel loss. In the following performance analysis, while calculating the energy efficiency of different RX architectures, power consumed by the clock generation circuits and biasing circuits has not been included for simplicity and to focus on the Rx architecture dependent power.

\subsection{Architecture I: Only sampler}
\begin{figure}[!]
\centering
\includegraphics[height=2in]{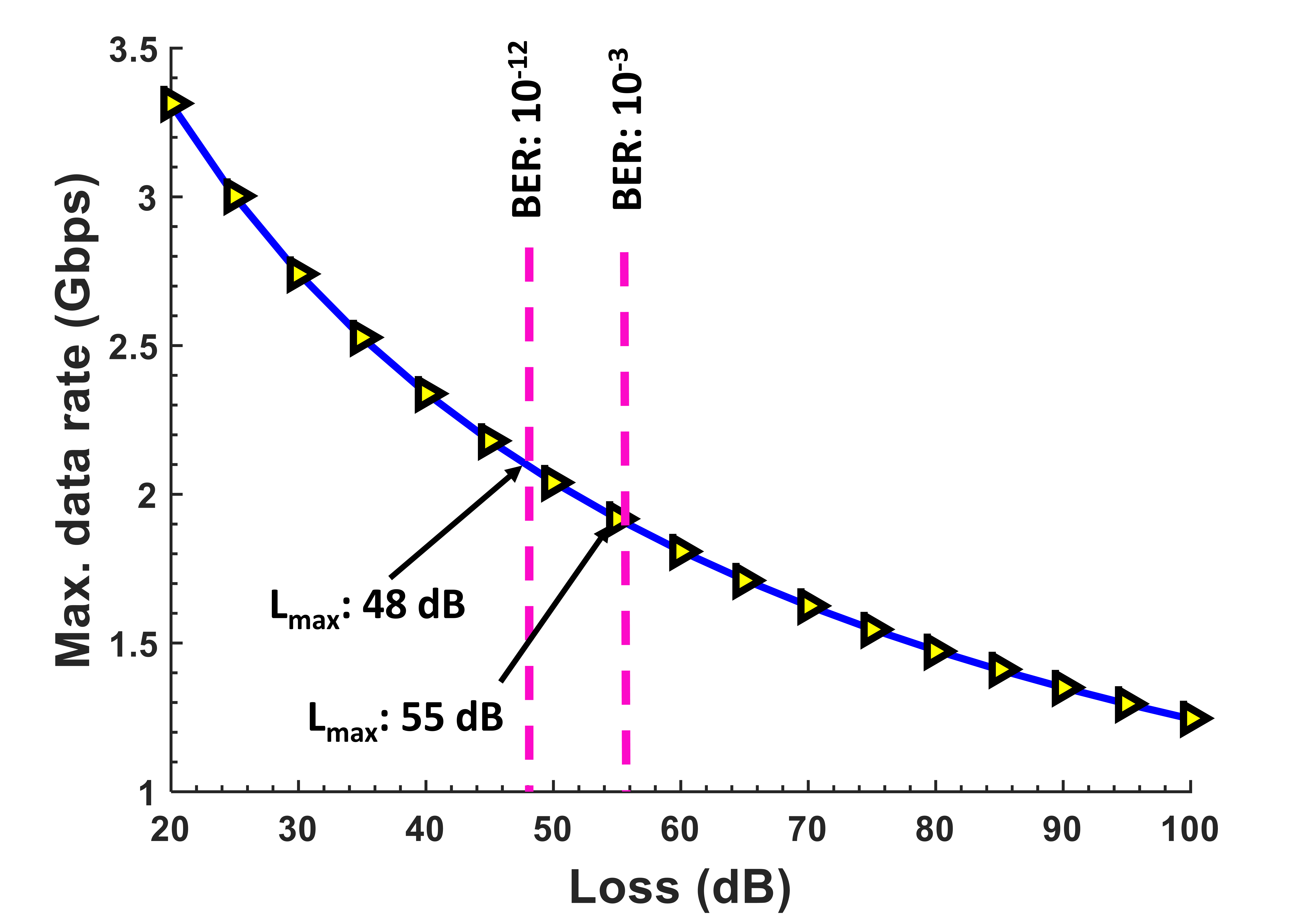}
\caption{Variation of maximum data rate with channel loss for architecture-1}
\label{top1_max_rate}
\end{figure}

Assuming a $1$ V transmitted signal swing, the input signal swing of the RX ($v_{RX})$ can be computed as a function of the channel loss, $L$ and maximum achievable data rate for that particular channel loss can be found following the methodology discussed earlier. Fig. \ref{top1_max_rate} shows the corresponding plot. Note that, $A_{FE}(f)=1$ for architecture-I. Also, as the channel loss increases, the input SNR reduces which degrades the BER of the received signal. Considering a target BER of $10^{-12}$, the maximum channel loss architecture-I can support is $48$ dB and for a BER of $10^{-3}$, the maximum allowable channel loss is $55$ dB. Also, as power consumption of the strongARM latch is proportional to the clock frequency (and hence to the data rate), energy efficiency of architecture-I is constant and independent of the data rate. For the design in this work the energy efficiency is found out to be $0.022$ pJ/bit.

\subsection{Architecture II: LNA + sampler}
\begin{figure}[!]
\centering
\includegraphics[height=2in]{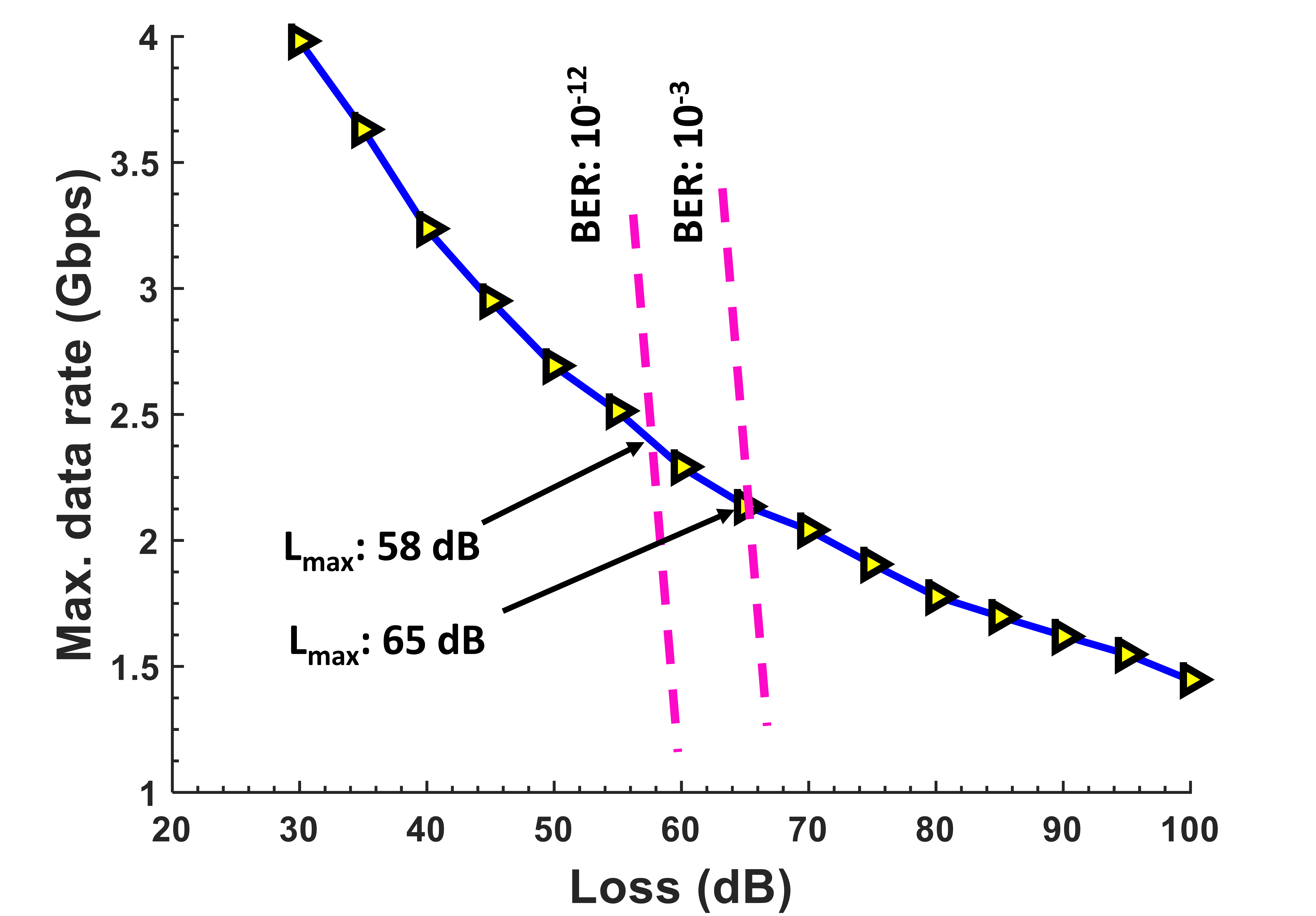}
\caption{Variation of maximum data rate with channel loss for architecture-II. Note that, deploying an LNA in the front-end enhances the data-rate by providing a gain to the input signal and extends the maximum allowable channel loss ($L_{max}$) for a target BER by reducing the total input-referred noise.}
\label{top2_max_rate}
\end{figure}

Fig. \ref{top2_max_rate} shows the performance of architecture-II for different channel loss. As expected, introduction of the LNA improves the maximum data rate for each channel loss and also shifts the BER constraint curve towards right by offering lower input referred noise. The input referred noise of the RX in this case can be given as
\begin{equation}
    V_{n,in,RX}=\sqrt{V_{n,in,LNA}^2+\frac{V_{n,in,samp}^2}{A_{LNA}^2}}
    \label{noise_top2}
\end{equation}

From eq. (\ref{noise_top2}) it can be seen that the RX input referred noise is mostly dominated by the LNA and hence is quite low comparative to that of architecture-I leading to a $10$ dB improvement in maximum allowable channel loss (Fig. \ref{top2_max_rate}). Also, the $V_{n,in,RX}$ is a function of data rate and hence, the BER constraint curves in Fig. \ref{top2_max_rate} are not exactly vertical as in Fig. \ref{top1_max_rate}. Hence, after $L_{max}$, the maximum data rate decreases rapidly while satisfying the BER constraint. Note that the minimum bias current required for the LNA varies linearly with data rate (Fig. \ref{curr_LNA}) rendering the energy efficiency of architecture-II to be constant which is having a value of $0.082$ pJ/bit in this design.

\subsection{Architecture III: Integrator + sampler}
\begin{figure}[!]
\centering
\includegraphics[height=2in]{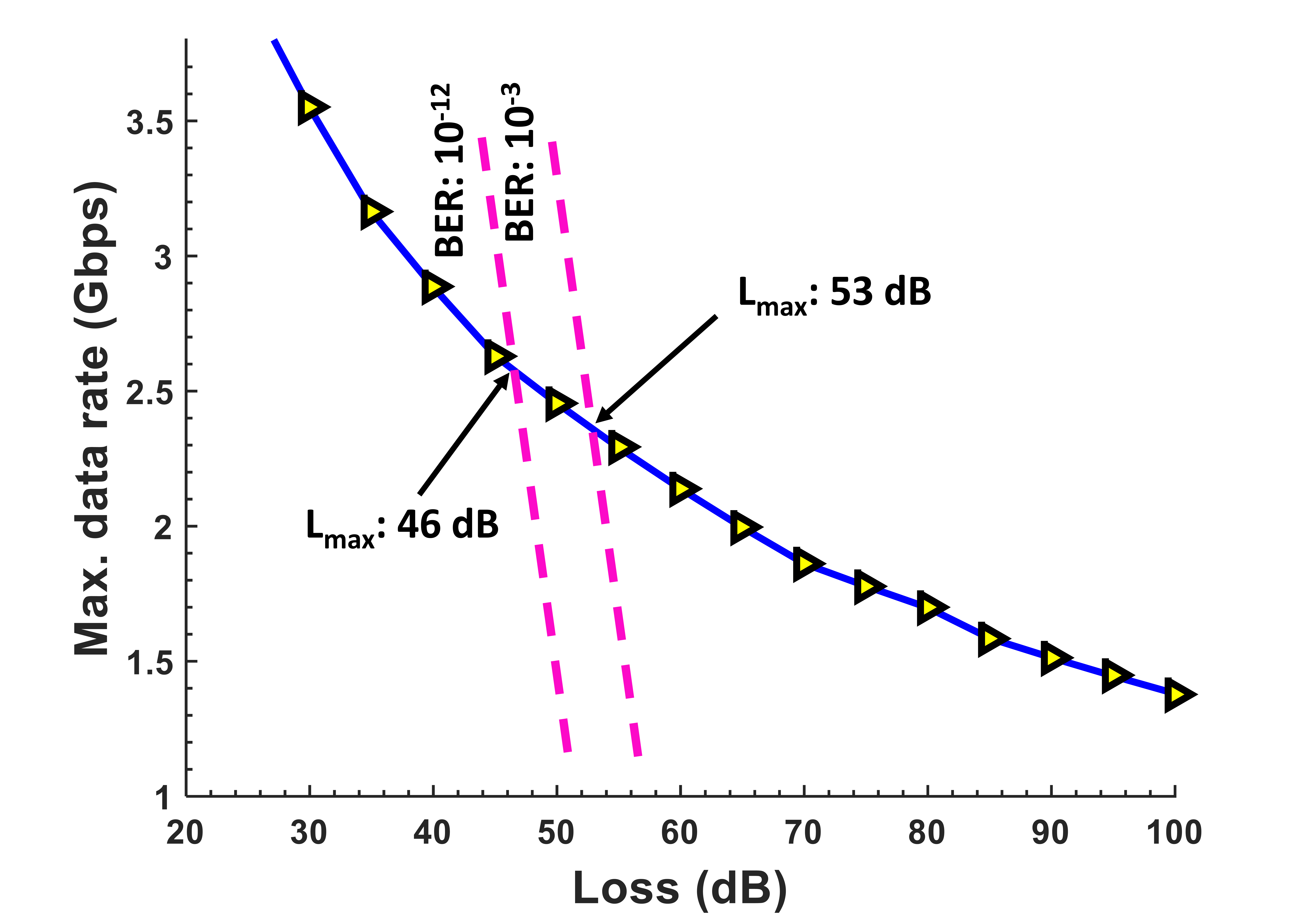}
\caption{Variation of maximum data rate with channel loss for architecture-III. Using integrator in the front-end improves the data-rate as compared to architecture-I, however, maximum allowable channel loss doesn't improve as input referred noise of an integrator is comparable or larger to that of a SAL.}
\label{top3_max_rate}
\end{figure}

Fig. \ref{top3_max_rate} shows the performance of architecture-III for different channel loss. Given that both the LNA in architecture-II and integrator in architecture-III are driving the same sampler (i.e. same $C_L$), performance of the integrator is subordinate to that of the LNA both in terms of gain and input referred noise. Hence, a deterioration in maximum allowable data rate and maximum achievable channel loss can be observed as compared to architecture-II. However, energy efficiency of this architecture ($\approx 0.042$ pJ/bit) is superior than that of architecture-II due to lower power consumption of the integrator. It also shows a significant improvement in the maximum data rate compared to that of architecture-I due to the additional gain provided by the integrator. Hence, architecture-III can be suitably used to achieve high data-rate with low power consumption for applications with relatively lower channel loss.

\subsection{Architecture IV: LNA + integrator(s)+ sampler}

\begin{figure}[!]
\centering
\includegraphics[height=2in]{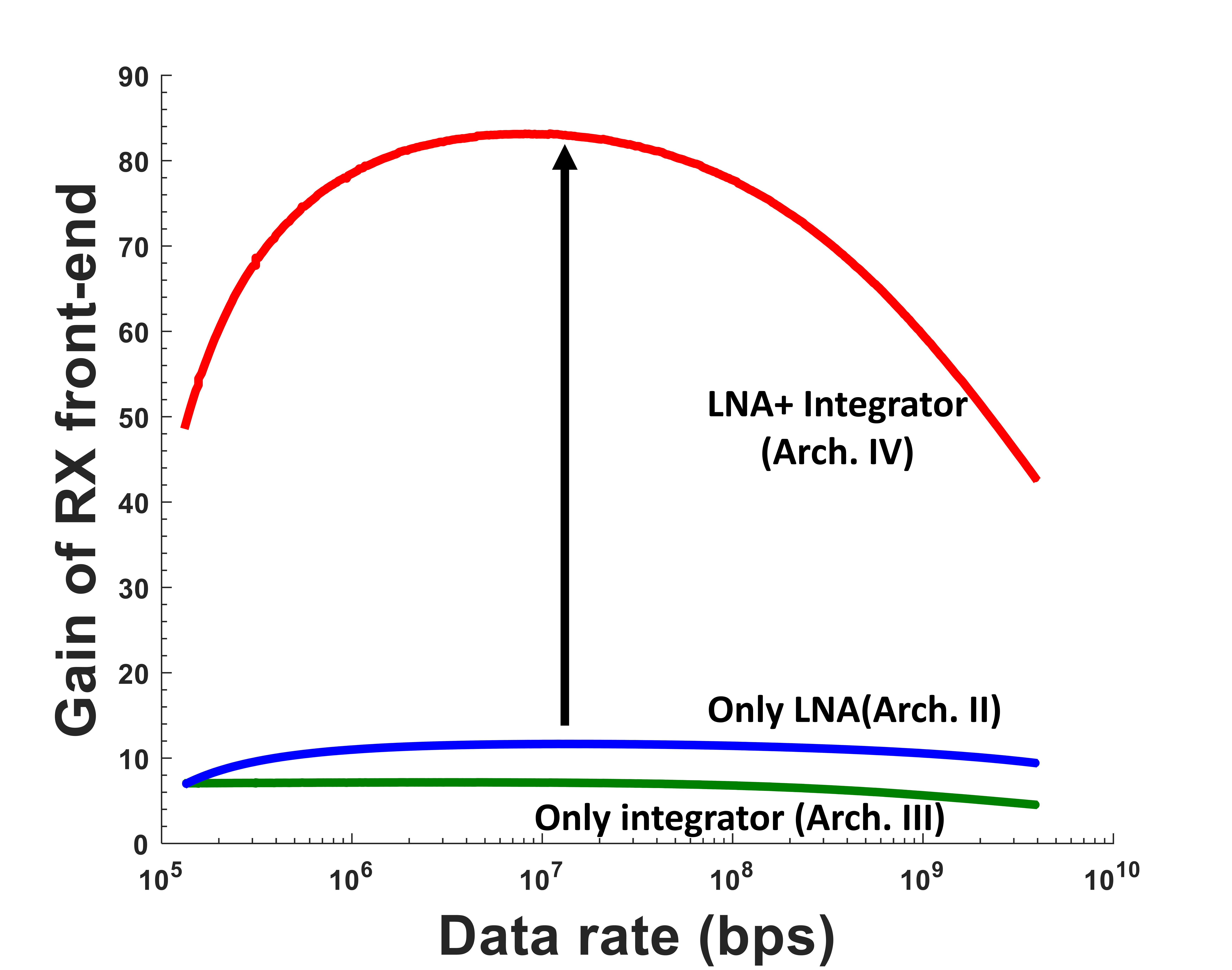}
\caption{Gain improvement of RX front-end by cascading LNA and integrator in architecture-IV.}
\label{gain_LNA_int}
\end{figure}

\begin{figure}[!]
\centering
\includegraphics[height=2in]{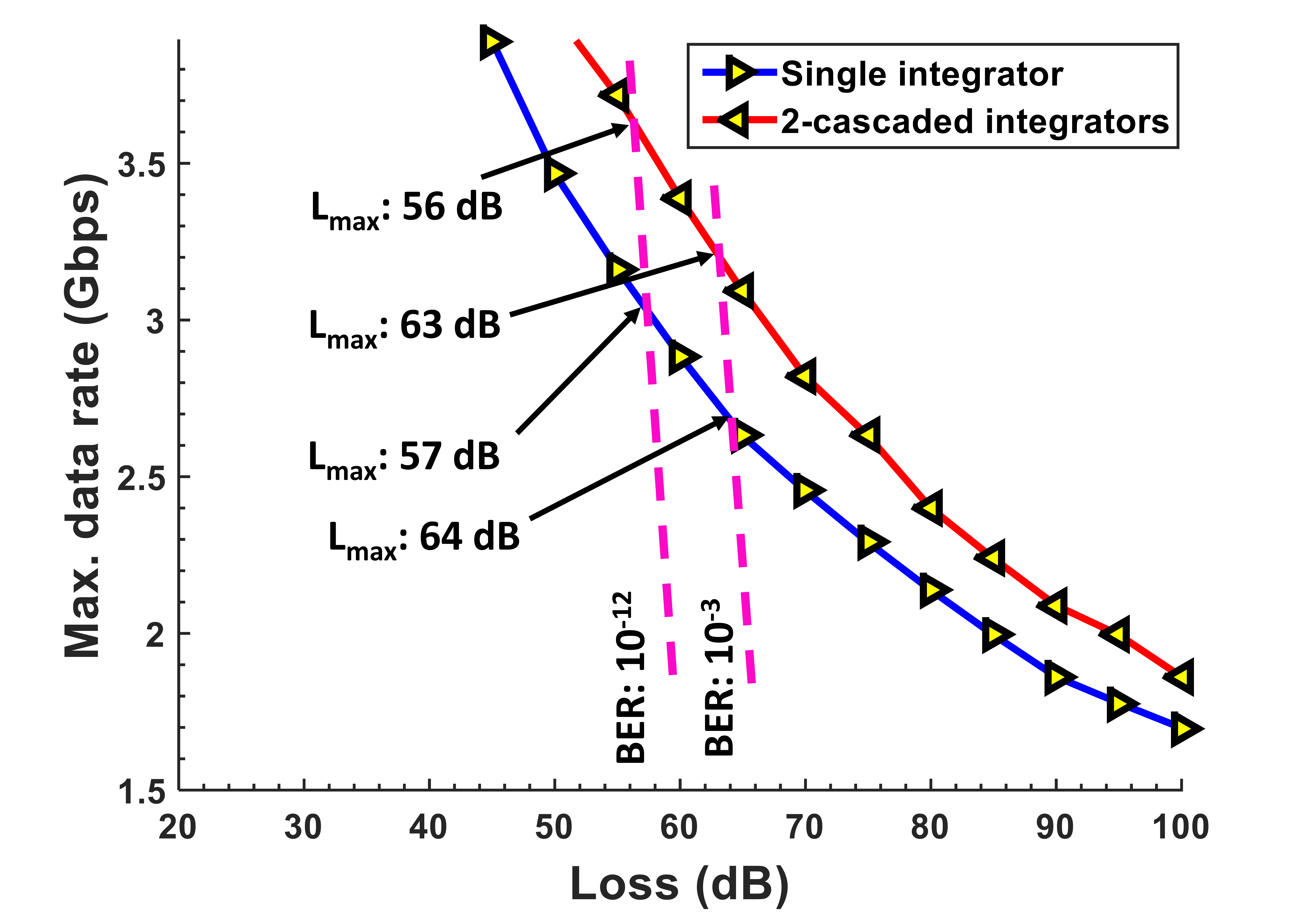}
\caption{Variation of maximum data rate with channel loss for architecture-IV with single and 2-cascaded integrators. Deploying an LNA followed by an integrator enhances the data-rate significantly as well as improves the maximum allowable channel loss by simultaneously providing large gain and reducing the input referred noise. Cascading two integrators provides even larger front-end gain and improves the data-rate further.}
\label{top4_max_rate}
\end{figure}

\begin{figure*}[!t]
\centering
\subfigure[]{
\includegraphics[height=2.2in]{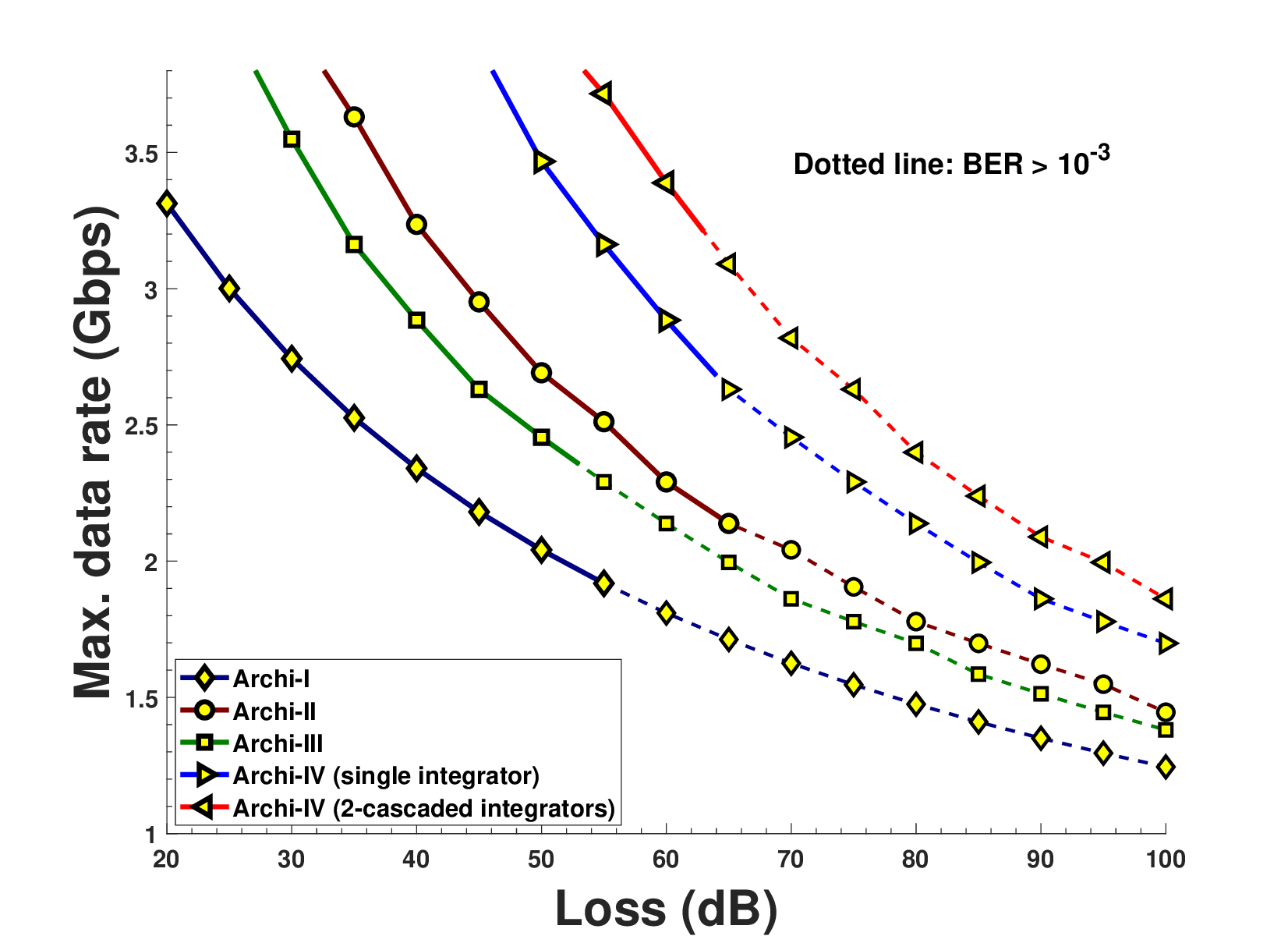}
\label{arch_data_rate_comp}
}
%\hspace*{2in}
\subfigure[]{
\includegraphics[height=2.2in]{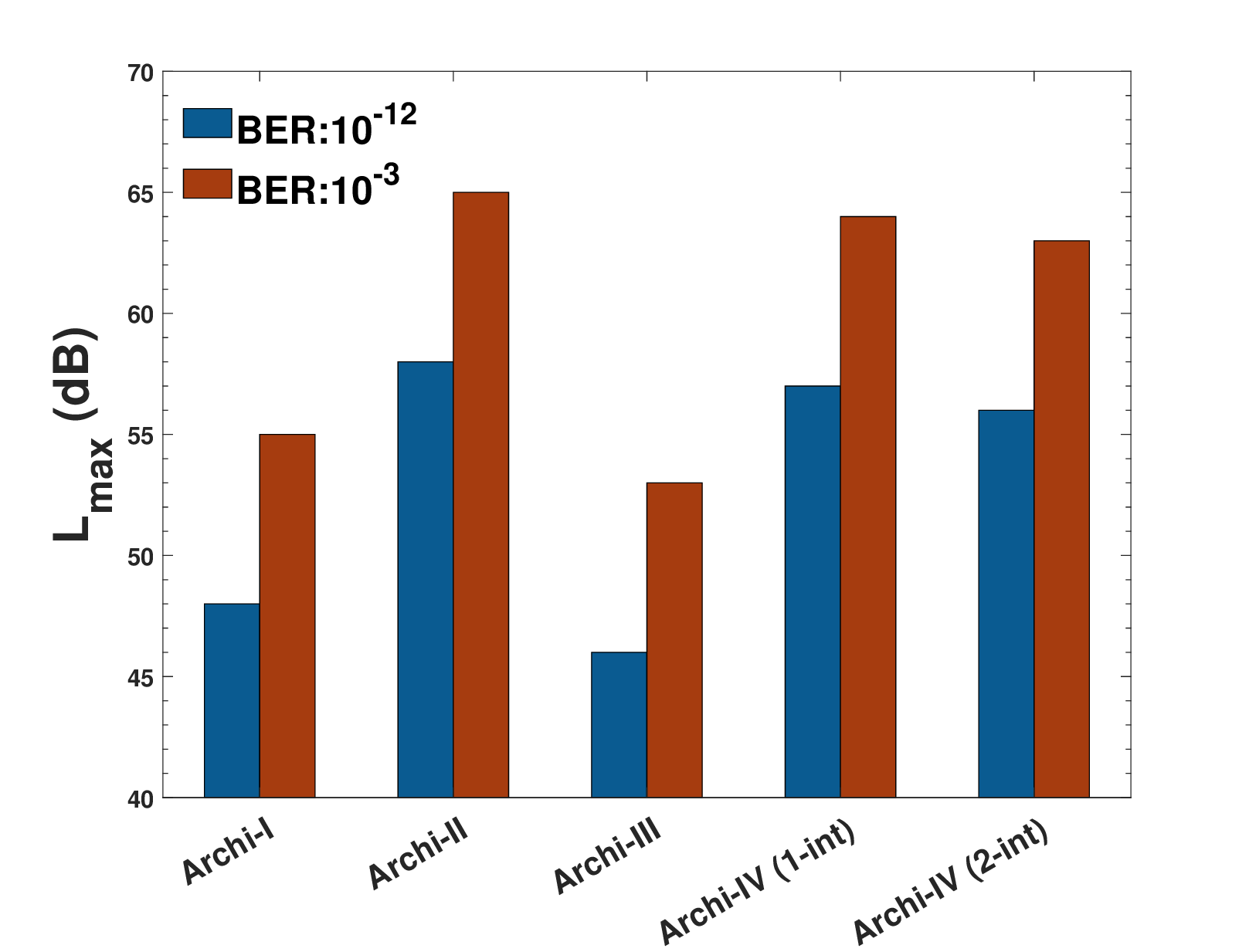}
\label{L_max_comp}
}
\hspace*{-0.3in}
\subfigure[]{
\includegraphics[height=2in]{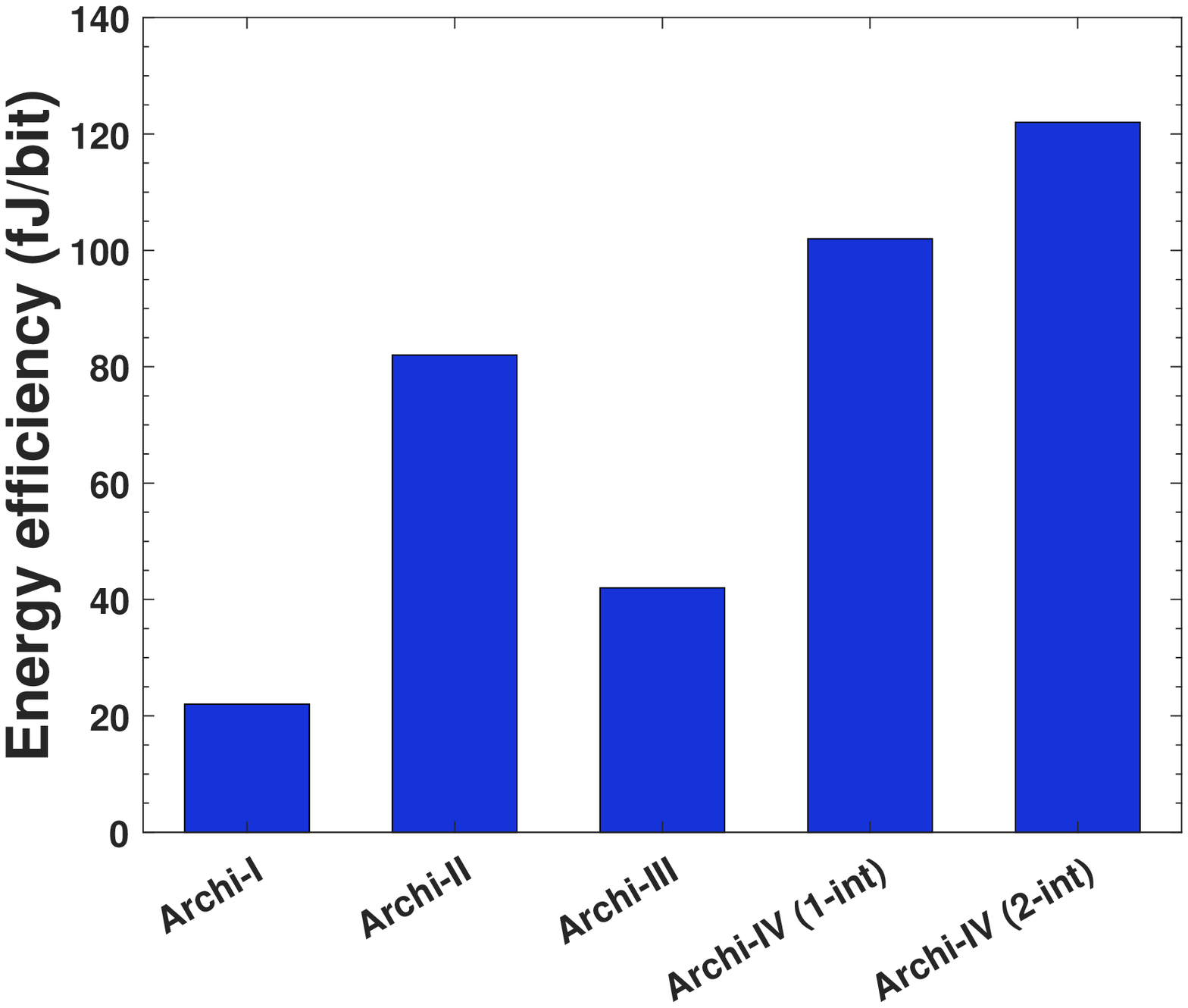}
\label{energy_eff_comp}
}

\caption{Performance comparison of different RX architectures in terms of (a) maximum achievable data rate for different channel loss, (b) maximum allowable channel loss for a target BER and (c) optimum energy efficiency}
\label{comparison_arch}
\end{figure*}

To improve the maximum achievable data rate further for architecture-II, an integrator can be introduced between the LNA and sampler, which provides gain to the signal with lower power consumption. Fig. \ref{gain_LNA_int} shows the improvement in gain of the RX front-end by cascading an LNA and integrator. Using both LNA and integrator in the RX front-end ensures both lower RX input referred noise($V_{n,in,RX}$) as well as higher front-end gain. Note that, an alternative way to improve the gain would have been to cascade multiple LNAs. However, the LNA$+$integrator combination provides comparable gain to that of LNA$+$LNA combination with a much lower power consumption. To increase the front-end gain further, multiple integrators can be used as discussed in section $IV.D$. Fig. \ref{top4_max_rate} shows the variation of maximum achievable data rate with channel loss for architecture-$IV$ with both single and 2-cascaded integrators. It can be seen from Fig. \ref{top4_max_rate} that the BER constraint curve remains the same for both single and 2-cascaded integrators. This is due to the fact that $V_{n,in,RX}$ is governed by the input referred noise of the LNA and first integrator as the large gain of the LNA and integrator combination makes the noise contribution of the next stages insignificant. Energy efficiency of architecture-$IV$ is $0.102$ pJ/bit with a single integrator and $0.122$ pJ/bit for 2-cascaded integrators.

\section{Comparison of different architectures}
Fig. \ref{comparison_arch} compares the performance of different RX architectures. It is obvious from the previous discussions that for any given channel loss, architecture-IV with 2-cascaded integrators gives the highest maximum achievable data-rate (Fig. \ref{arch_data_rate_comp}). With increase in channel loss, the maximum achievable data-rate decreases with a gradual degradation in the BER performance for all the architectures. In Fig. \ref{arch_data_rate_comp}, the dotted portion of the curves represents a BER $>10^{-3}$. Fig. \ref{L_max_comp} shows the maximum sustainable channel loss for different RX architectures for two standard BERs. It can be seen that deploying LNA in the RX front-end extends the maximum achievable channel loss for architectures II and IV compared to the others. Hence, if the application demands high data rate even with a very large channel loss, one must deploy an LNA in the RX front-end. Note that, all these figures show the ideal performance of different architectures. In real scenario, the clocking scheme apart from the signaling blocks may also limit the maximum data-rate and increase the energy-efficiency. However, those additional constrains have not been considered in this work as the main focus was to identify the most suitable architecture given different channel loss profile and data-rate requirements. Our previous work on interference-robust HBC (\cite{HBC_JSSC}) deploys architecture-III which achieves a data-rate of 30 Mbps for a channel loss of 60 dB and target BER of $10^{-3}$. This observation can be corroborated from Fig. \ref{top3_max_rate} which shows that, for architecture-III once the channel loss exceeds $L_{max}$ (=$53$ dB), the maximum achievable data-rate decreases drastically while satisfying the target BER.

\section{conclusion}

The work theoretically analyzes performance of different signaling blocks that can be employed in the RX for broad-band communication through lossy wireline-like channels. A new approach to theoretically estimate the input referred noise of the strongARM latch has been described and compared with simulation results. An accurate closed-form expression of the gain of the current-integrating amplifier has been developed. The work also proposes the use of multi-integrator cascade as a low-power gain element and shows how employing the same in the RX improves the gain of the front-end with low power consumption. Finally, based on the analysis of the signaling blocks, performance of different architectures have been analyzed and compared. The in-depth analysis sets a foundation for the choice of appropriate receiver architecture for lossy broadband channels, that are becoming popular in applications such as proximity communication, human body communication among others. 

\section{acknowledgement}
This work is supported in parts by the Semiconductor Research Corporation (SRC) under Grant 2878.014 and the National Science Foundation (NSF) CAREER Award under Grant ECCS 1944602.

\end{document}